\documentclass[a4paper,11pt]{article}
\usepackage{jheppub}
\usepackage[T1]{fontenc}
\usepackage[utf8]{inputenc}
\usepackage{amsmath}
\usepackage{amssymb}
\usepackage{bm}

\begin{document}
\title{Feynman Integrals from Positivity Constraints}
\author[1]{Mao Zeng}
\affiliation[1]{Higgs Centre for Theoretical Physics, University of Edinburgh,\\
James Clerk Maxwell Building, Peter Guthrie Tait Road, Edinburgh, EH9 3FD,\\
United Kingdom}
\emailAdd{mao.zeng@ed.ac.uk}
\abstract{
  We explore inequality constraints as a new tool for numerically evaluating Feynman integrals. A convergent Feynman integral is non-negative if the integrand is non-negative in either loop momentum space or Feynman parameter space. Applying various identities, all such integrals can be reduced to linear sums of a small set of master integrals, leading to infinitely many linear constraints on the values of the master integrals. The constraints can be solved as a semidefinite programming problem in mathematical optimization, producing rigorous two-sided bounds for the integrals which are observed to converge rapidly as more constraints are included, enabling high-precision determination of the integrals. Positivity constraints can also be formulated for the $\epsilon$ expansion terms in dimensional regularization and reveal hidden consistency relations between terms at different orders in $\epsilon$. We introduce the main methods using one-loop bubble integrals, then present a nontrivial example of three-loop banana integrals with unequal masses, where 11 top-level master integrals are evaluated to high precision.

}

\maketitle
\tableofcontents

\section{Introduction}
Evaluation of Feynman integrals is both a classic problem and an expanding frontier of research in quantum field theory, with many areas of applications such as collider physics, statistical mechanics, and gravitational physics. A widely used method for numerical evaluations of Feynman integrals is Monte Carlo integration with sector decomposition \cite{Binoth:2000ps, Bogner:2007cr, Kaneko:2009qx, Borowka:2015mxa, Borowka:2017idc, Smirnov:2013eza, Smirnov:2015mct}. Another method, numerical series solutions of differential equations \cite{Moriello:2019yhu, Hidding:2020ytt, Liu:2017jxz, Liu:2021wks, Liu:2022chg, Armadillo:2022ugh, Hidding:2022ycg}, enjoyed intense developments in the last few years. Some other methods, which are either inherently numerical or can be used numerically, include Mellin-Barnes representations \cite{Usyukina:1992jd, Usyukina:1993ch, Smirnov:1999gc, Tausk:1999vh, Czakon:2005rk, Smirnov:2009up, Gluza:2007rt, Belitsky:2022gba}, difference equations from dimensional recursion \cite{Laporta:2000dsw, Lee:2012te, Lee:2022art}, Loop-tree duality \cite{Catani:2008xa, Runkel:2019yrs, Capatti:2019ypt}, and tropical Monte Carlo integration \cite{Borinsky:2020rqs, Borinsky:2023jdv}. See also books \cite{Weinzierl:2022eaz, Smirnov:2012gma} which comprehensively review both numerical and analytic methods. Despite these developments, numerical evaluation of Feynman integrals still presents challenges in the ongoing quest for higher precisions in perturbative calculations, and new explorations are warranted.

A fruitful recent idea in theoretical physics is the use of positivity constraints, e.g.\ arising from the unitarity of a Hilbert space, to constrain unknown parameters from first principles, sometimes reaching great accuracy. Prominent examples include the conformal bootstrap \cite{Rattazzi:2008pe}, the non-perturbative S-matrix bootstrap (see review \cite{Kruczenski:2022lot} and references within), and EFT positivity bounds \cite{Adams:2006sv, Bellazzini:2020cot, Caron-Huot:2020cmc, Tolley:2020gtv, Arkani-Hamed:2020blm, Caron-Huot:2021rmr, Caron-Huot:2022ugt}. Some of the predictions in the S-matrix and EFT contexts have been checked against explicit perturbative calculations involving Feynman integrals \cite{Bern:2021ppb, Chen:2022nym, Bellazzini:2022wzv}, so it is natural to explore positivity properties for Feynman integrals themselves. What directly inspired this paper, though, is recent applications of positivity constraints to bootstrapping simple quantum mechanical systems and matrix models \cite{Lin:2020mme, Han:2020bkb, Berenstein:2021dyf, Kazakov:2021lel, Berenstein:2022unr} as well as lattice models \cite{Anderson:2016rcw, Kazakov:2022xuh, Cho:2022lcj}, which crucially use various linear identities that are reminiscent of integration-by-parts identities for Feynman integrals \cite{Chetyrkin:1981qh} as well as a numerical technique known as semidefinite programming \cite{vandenberghe1996semidefinite} which will be adapted to our calculations. Semidefinite programming was introduced to theoretical physics by Ref.~\cite{Poland:2011ey} in the conformal bootstrap and subsequently applied to wider contexts.

In this work, we will formulate positivity constraints for \emph{Euclidean Feynman integrals}, which can be either (1) integrals in Euclidean spacetime or can be rewritten in Euclidean spacetime after a trivial Wick rotation, or (2) integrals in Minkowskian spacetime but with kinematics in the so called \emph{Euclidean region}, i.e.\ with center-of-mass energy of incoming momenta below the particle production threshold. In case (1), the loop integrand in Euclidean momentum space has non-negative propagator denominators, and the integrand is non-negative as long as the numerator is non-negative. In case (2), the integral is real due to the absence of Cutkosky cuts. After Feynman parametrization, the parametric integrand involves non-negative graph polynomials and stays non-negative when multiplied by a positive function of the Feynman parameters. In both cases, a non-negative integrand implies a non-negative integral as long as the integral is convergent, i.e.\ has no ultraviolet or infrared divergences.
The initial restriction to convergent integrals is not a fundamental limitation, as divergent integrals can be rewritten as linear sums of convergent integrals multiplied by divergent coefficients \cite{vonManteuffel:2014qoa}, and then it suffices to evaluate the convergent integrals as a Taylor series in the dimensional regularization parameter $\epsilon$.

We will see that positivity constraints, expressed in the language of semidefinite programming, are strong enough to precisely determine the values of Feynman integrals. Moreover, rigorous error bounds can be obtained. Our machinery relies on linear relations between Feynman integrals, including integration-by-parts identities \cite{Chetyrkin:1981qh} and dimension-shifting identities \cite{Tarasov:1996br}, to express all positivity constraints as linear constraints on the values of a set of master integrals.

The outline of the paper is as follows. In Section \ref{sec:bubble}, we introduce the main methods using the simple example of massive one-loop bubble integrals. Specifically, Subsection \ref{subsec:bubbleIBP} gives a short review of two kinds of linear identities for Feynman integrals, integration-by-parts identities and dimension shifting identities. Subsection \ref{subsec:posMomSpace} discusses positivity constraints in Euclidean momentum space, starting from ad hoc constraints that narrow down the allowed value of the bubble master integral to $\sim 50\%$ accuracy at an example kinematic point, then developing the machinery of semidefinite programming to reach an accuracy of $10^{-14}$. We switch to Feynman parameter space in Subsection \ref{subsec:posFeynSpace}, which allows calculating the integrals in a wider region of kinematic parameters below the two-particle cut threshold, while many steps of the calculations are unchanged from the momentum-space case. Subsection \ref{subsec:epExpansion} formulates positivity constraints for $\epsilon$ expansions of Feynman integrals in dimensional regularization. Subsection \ref{subsec:numericalDiff} presents an alternative method for calculating the $\epsilon$ expansion based on numerical differentiation of results w.r.t.\ the spacetime dimension. Having laid out most of the methods, in Section \ref{sec:banana}, we present a nontrivial application to three-loop banana integrals with four unequal masses. Due to a large number of undetermined master integrals, semidefinite programming becomes essential in efficiently solving the positivity constraints. We calculate all master integrals at an example kinematic point up to the second order in $\epsilon$ expansion in $d=2-2\epsilon$, and a detailed account of the numerical accuracies is given. We end with some discussions in Section \ref{sec:discussions}.

\section{One-loop bubble integrals}
\label{sec:bubble}
\begin{figure}
  \centering
  \includegraphics[width=0.3\textwidth]{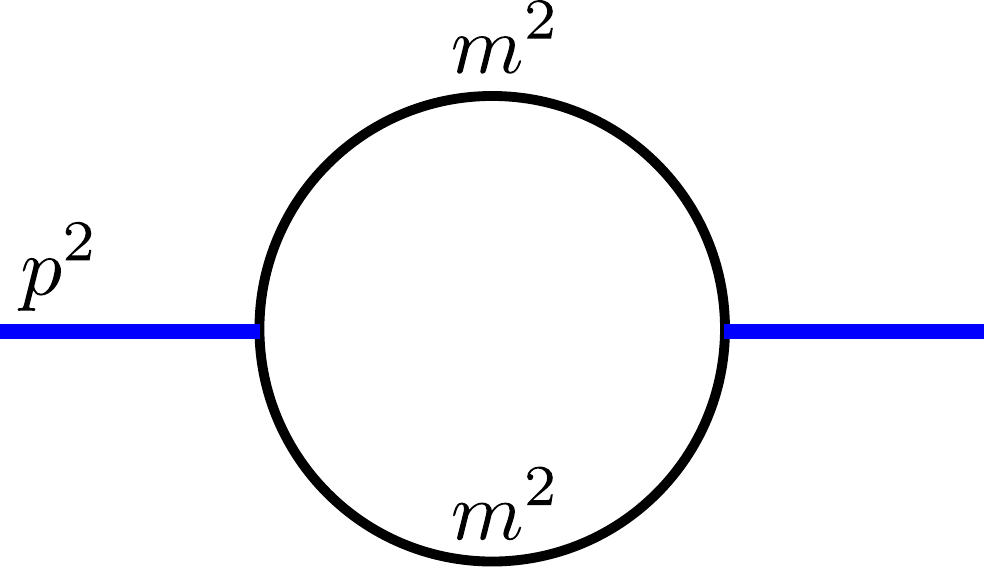}
  \caption{The one-loop bubble integral with external legs of virtuality $p^2$ and two internal massive line with the same squared mass $m^2$.}
  \label{fig:bubble}
\end{figure}

We consider the following family of Feynman integrals in Minkowski spacetime parametrized by two integers $a_1$ and $a_2$,
\begin{equation}
    I_{a_1,a_2}^d \equiv \int \frac {d^d l \, e^{\gamma_E \epsilon}}{i \pi^{d/2}} \frac{1} {(- l^2 + m^2)^{a_1} [-(p + l)^2 + m^2]^{a_2}} , \label{eq:selfEnergyInt}
\end{equation}
which correspond to the diagram in Fig.~\ref{fig:bubble} but with the propagator denominators raised to general integer powers.
The external mass is $\sqrt{p^2}$ and the internal mass for the two propagators is $m$, with the two propagators raised to powers $a_1$ and $a_2$. If either $a_1$ or $a_2$ is non-positive, Eq.~\eqref{eq:selfEnergyInt} becomes a massive tadpole integral possibly with a numerator. If $a_1$ and $a_2$ are both non-positive, the integral vanishes in dimensional regularization. The spacetime dimension $d$ is equal to $4- 2 \epsilon$, with $\epsilon$ being the dimensional regularization parameter.
Eq.~\eqref{eq:selfEnergyInt} can be re-written, by Wick rotation of the integration contour, as the following integrals in \emph{Euclidean} spacetime,
\begin{equation}
    I_{a_1,a_2}^d \equiv \int \frac {d^d \bm l \, e^{\gamma_E \epsilon}}{\pi^{d/2}} \frac{1} {(\bm l^2 + m^2)^{a_1} [(\bm p + \bm l)^2 + m^2]^{a_2}} , \label{eq:selfEnergyIntEuc} \, .
\end{equation}
where $\bm p^2 \equiv -p^2$.

We will use positivity constraints to numerically evaluate bubble integrals Eq.~\eqref{eq:selfEnergyInt} for $p^2 < 4m^2$, i.e.~below the two-particle production threshold. We first consider the case $p^2 < 0$, i.e.~$\bm p^2 > 0$. In this case, $\bm p$ can be literally embedded in Euclidean spacetime as a vector with real-valued components, and we will derive positivity constraints starting from the loop momentum integral Eq.~\eqref{eq:selfEnergyIntEuc}. We will subsequently present a treatment applicable to all $p^2 < 4m^2$ by using Feynman parameter representations of the integrals. Before actual calculations, we first briefly review linear identities for the Feynman integrals involved, arising from \emph{integration by parts} and \emph{dimension shifting}. Efficiently solving IBP identities for more complicated Feynman integrals is a major research problem, and we refer readers to Refs.~\cite{Laporta:2000dsw, Laporta:1996mq, Anastasiou:2004vj, vonManteuffel:2012np, Smirnov:2008iw, Smirnov:2019qkx, Lee:2012cn, Lee:2013mka, Maierhofer:2017gsa} for relevant computational algorithms and software.

\subsection{Review: integration-by-parts (IBP) and dimensional-shifting identities}
\label{subsec:bubbleIBP}
$I^d_{a_1,a_2}$ at different values of $a_1$ and $a_2$, as defined in Eq.~\eqref{eq:selfEnergyInt}, are related through integration-by-parts (IBP) identities \cite{Chetyrkin:1981qh}, as total derivatives integrate to zero without boundary terms in dimensional regularization. To derive the identities, we will write dot products as linear combinations of denominators and constants,
\begin{equation}
  l^2 = -(-l^2 + m^2) + m^2, \qquad 2 p \cdot l = -[-(p + l)^2 + m^2] + [- l^2 + m^2] - p^2 \, .
\end{equation}
The actual identities are
\begin{align}
  & \int \frac {d^d l \, e^{\gamma_E \epsilon}}{i \pi^{d/2}} \frac{\partial}{\partial l^\mu} \frac{p^\mu} {(- l^2 + m^2)^{a_1} [-(p + l)^2 + m^2]^{a_2}} = 0 \nonumber \\
  & \implies -a_1 p^2 I^d_{a_1+1,a_2} + a_2 p^2 I^d_{a_1,a_2+1} - a_1 I^d_{a_1+1,a_2-1} + (a_1-a_2) I^d_{a_1,a_2} + a_2 I^d_{a_1-1,a_2+1} = 0 \, , \label{eq:bubbleIBP1}
\end{align}
and
\begin{align}
  & \int \frac {d^d l \, e^{\gamma_E \epsilon}}{\pi^{d/2}} \frac{\partial}{\partial l^\mu} \frac{l^\mu} {(-l^2 + m^2)^{a_1} [-(p + l)^2 + m^2]^{a_2}} = 0 \implies \nonumber \\
  & 2a_1 m^2 I^d_{a_1+1,a_2} + a_2 (-p^2 + 2m^2) I^d_{a_1,a_2+1} + (d- 2a_1 - a_2) I^d_{a_1,a_2} - a_2 I^d_{a_1-1,a_2+1} = 0 \, , \label{eq:bubbleIBP2}
\end{align}
We also need the diagram reflection symmetry relations from relabeling $l \rightarrow -(p + l)$,
\begin{equation}
  I^d_{a_1, a_2} = I^d_{a_2, a_1} \, , \label{eq:bubbleSym}
\end{equation}
and the boundary condition in dimensional regularization,
\begin{equation}
  I^d_{a_1, a_2} = 0, \ \text{when both } a_1 \leq 0 \ \text{and }  a_2 \leq 0 \, . \label{eq:bubbleRestriction}
\end{equation}
Solving eqs.~\eqref{eq:bubbleIBP1}-\eqref{eq:bubbleRestriction}, e.g.\ by iteratively eliminating $I^d_{a_1,a_2}$ with the largest values of $|a_1| + |a_2|$, allow us to rewrite any $I^d_{a_1, a_2}$ with fixed spacetime dimension $d$ in terms of two \emph{master integrals},
\begin{equation}
  I^d_{1,1}, \quad I^d_{1,0}, \label{eq:bubbleNaiveBasis}
\end{equation}
i.e.\ a bubble integral and a tadpole integral. We will then change to an alternative basis $I_1$ and $I_2$, which is ultraviolet finite as $d$ approaches $4$ and in fact for any $d<6$,
\begin{equation}
  \begin{aligned}
    I^d_{2,1} &= \frac {d-3} {p^2 - 4 m^2} I^d_{1,1} + \frac {d-2} {2 m^2 (p^2 - 4 m^2)} I^d_{1,0} \, , \\
    I^d_{3,0} &= \frac {(d-2)(d-4)}{8 m^4} I^d_{1,0} \, .
  \end{aligned}
  \label{eq:bubbleFiniteBasis}
\end{equation}
\emph{For the rest of this paper, we will impose $d<6$ in the treatment of the bubble integrals and pay special attention to values of $d$ near $4$.}

The process of carrying out the above calculation and rewriting any integral as linear sums of master integrals is referred to as \emph{IBP reduction}. Here are two examples showing how other integrals are expressed as linear combinations of the basis Eq.~\eqref{eq:bubbleFiniteBasis},
\begin{align}
  I^d_{2,2} &= \frac{1}{p^2 (4m^2-p^2)} \Big( [(6-d)p^2 - 4m^2] I^d_{2,1} + 4 m^2 I^d_{3,0} \Big) \label{eq:I22result} \\
  I^d_{3,1} &= I^d_{1,3} = \frac{1}{2p^2 (4m^2-p^2)} \Big( [(4-d)p^2 + 4m^2] I^d_{2,1} + 2(p^2-2m^2) I^d_{3,0} \Big) \, . \label{eq:I31result}
\end{align}
The tadpole integral $I^d_{3,0}$ is well known in textbooks. The bubble integral with one of the propagator raised to a higher power, $I^d_{2,1}$, is also known analytically, but we will use positivity constraints to numerically evaluate it for the purpose of demonstrating our methods. We will write
\begin{align}
  I^d_{2,1} = \hat I^d_{2,1} \cdot I^d_{3,0} \, ,
\end{align}
and treat the normalized bubble integral
\begin{equation}
  \hat I^d_{2,1} = I^d_{2,1} / I^d_{3,0} \label{eq:bubbleMasterParam}
\end{equation}
as an unknown quantity to be bounded by positivity constraints. More generally, we define a ``hatted'' notation for integrals normalized against the finite tadpole integral,
\begin{equation}
  \hat I^d_{a_1, a_2} = I^d_{a_1, a_2} / I^d_{3,0} , \, \label{eq:bubbleNormalizedA1A2}
\end{equation}
under which Eqs.~\eqref{eq:I22result} and \eqref{eq:I31result} become
\begin{align}
  \hat I^d_{2,2} &= \frac{1}{p^2 (4m^2-p^2)} \Big( [(6-d)p^2 - 4m^2] \hat I^d_{2,1} + 4 m^2 \Big) \label{eq:I22resultNormalized} \\
  \hat I^d_{3,1} &= I^d_{1,3} = \frac{1}{2p^2 (4m^2-p^2)} \Big( [(4-d)p^2 + 4m^2] \hat I^d_{2,1} + 2(p^2-2m^2) \Big) \, . \label{eq:I31resultNormalized}
\end{align}
We also need dimensional-shifting identities.
Using the Schwinger parametrization, Eq.~\eqref{eq:selfEnergyInt} is rewritten as
\begin{equation}
  I^d_{a_1, a_2} = \frac {e^{\gamma_E \epsilon}} {\Gamma(a_1) \Gamma(a_2)} \int_0^\infty d x_1 \int_0^\infty d x_2 \, x_1^{a_1-1} x_2^{a_2-1} (x_1 + x_2)^{-d/2} \exp (- i \mathcal U / \mathcal F) \, ,
  \label{eq:bubbleSchwinger}
\end{equation}
where $\mathcal U$ and $\mathcal F$ are graph polynomials depending on $x_1$ and $x_2$,
\begin{equation}
  \mathcal U(x_1, x_2) = x_1+x_2, \quad \mathcal F(x_1, x_2) = m^2(x_1+x_2)^2 - p^2 x_1 x_2 - i 0^+ \, .
  \label{eq:bubbleFpolySchwinger}
\end{equation}
Therefore, $(d-2)$ dimensional integrals can be written as
\begin{align}
  I^{d-2}_{a_1, a_2} &= \frac {e^{\gamma_E \epsilon}} {\Gamma(a_1) \Gamma(a_2)} \int_0^\infty d x_1 \int_0^\infty d x_2 \, x_1^{a_1-1} x_2^{a_2-1} (x_1 + x_2)^{-(d-2)/2} \exp (- i U / F) \nonumber \\
  &= \frac {e^{\gamma_E \epsilon}} {\Gamma(a_1) \Gamma(a_2)} \int_0^\infty d x_1 \int_0^\infty d x_2 \, x_1^{a_1-1} x_2^{a_2-1} (x_1 + x_2) (x_1 + x_2)^{-d/2} \exp (- i U / F) \nonumber \\
  &= \frac {e^{\gamma_E \epsilon}} {\Gamma(a_1) \Gamma(a_2)} \int_0^\infty d x_1 \int_0^\infty d x_2 \, \big( x_1^{a_1} x_2^{a_2-1} + x_1^{a_1-1} x_2^{a_2} \big)
  (x_1 + x_2)^{-d/2} \exp (- i U / F) \nonumber \\
  &= a_1 I^d_{a_1+1, a_2} + a_2 I^d_{a_1, a_2+1} \, ,
  \label{eq:bubbleDimShift}
\end{align}
where the last line used Eq.~\eqref{eq:bubbleFpolySchwinger} to map different monomials in $x_1$ and $x_2$ to bubble integrals with different propagator powers.
Applying Eq.~\eqref{eq:bubbleDimShift} to the finite bubble master integral $I^d_{2,1}$ on the LHS of Eq.~\eqref{eq:bubbleFiniteBasis} in $(d-2)$ spacetime dimensions and then performing IBP reduction, we obtain
\begin{equation}
  I^{d-2}_{2,1} = \frac {2 (d-5)} {p^2 - 4m^2} I^d_{2,1} - \frac 2 {p^2 - 4m^2} I^d_{3,0} \, . \label{eq:dimshift1}
\end{equation}
The dimension-shifting formula for the tadpole integral can be obtained easily, e.g.\ by using closed-form results for tadpole integrals. The result is
\begin{equation}
  I^{d-2}_{3,0} = \frac{6-d}{2 m^2} I^d_{3,0} \, . \label{eq:dimshift2}
\end{equation}
Eqs.~\eqref{eq:dimshift1} and \eqref{eq:dimshift2} are the dimension-shifting identities that express the two master integrals in Eq.~\eqref{eq:bubbleFiniteBasis} in lower spacetime dimensions to the same master integrals in higher spacetime dimensions. By inverting Eqs.~\eqref{eq:dimshift1} and \eqref{eq:dimshift2}, we obtain dimension shifting identities in the reverse direction, i.e.\ expressing master integrals in higher spacetime dimensions in terms of master integrals in lower spacetime dimensions,
\begin{align}
  I^{d+2}_{2,1} &= \frac {p^2 - 4m^2} {2(d-3)} I^d_{2,1} - \frac {2m^2} {(d-3)(d-4)} I^d_{3,0} \, , \label{eq:dimshift1prime} \\
  I^{d+2}_{3,0} &= \frac{2 m^2} {4-d} I^d_{3,0} \, . \label{eq:dimshift2prime}
\end{align}

\subsection{Positivity constraints in loop momentum space}
\label{subsec:posMomSpace}
Here we focus on the case $p^2 <0$ and use the Wick-rotated expression for the integrals, Eq.~\eqref{eq:selfEnergyIntEuc}, with
\begin{equation}
  \bm p^2 = M^2 = -p^2 \, .
\end{equation}
We will later present a Feynman parameter-space treatment that seems more powerful and in particular works for all $p^2 < 4m^2$, but the loop momentum space treatment here will help build up intuitions.
\subsubsection{A crude first attempt}
\label{subsubsec:crude}
We first consider the following two convergent integrals with non-negative integrands in loop momentum space,
\begin{equation}
  \begin{aligned}
    I^d_{2,2} &= \int \frac {d^d \bm l \, e^{\gamma_E \epsilon}}{\pi^{d/2}} \frac{1} {(\bm l^2 + m^2)^2 [(\bm p + \bm l)^2 + m^2]^2} \, , \\
    I^d_{3,1} + I^d_{1,3} - 2 I^d_{2,2}  &= \int \frac {d^d \bm l \, e^{\gamma_E \epsilon}}{\pi^{d/2}} \frac{1} {(\bm l^2 + m^2) [(\bm p + \bm l)^2 + m^2]} \\
    &\quad \times \left(  \frac 1 {\bm l^2 + m^2} - \frac 1 {(\bm p + \bm l)^2 + m^2 } \right)^2 \, .
  \end{aligned}
\end{equation}
We have used the notation Eq.~\eqref{eq:selfEnergyInt} and the subsequent Wick rotation Eq.~\eqref{eq:selfEnergyIntEuc}. Since the integrals are massive and have no infrared divergence, a simple ultraviolet power-counting shows that the above integrals are convergent near $4$ dimensions. These integrals therefore have finite non-negative values,
\begin{equation}
  I^d_{2,2} \geq 0, \qquad I^d_{3,1} + I^d_{1,3} - 2 I^d_{2,2} \geq 0 \, .
\end{equation}
By IBP reduction as described in Section \ref{subsec:bubbleIBP}, the above inequalities translate into constraints on the finite master integrals on the LHS of Eq.~\eqref{eq:bubbleFiniteBasis}. The needed IBP reduction results are shown in Eqs.~\eqref{eq:I22result} and \eqref{eq:I31result}, in which we will rewrite $p^2=-M^2$. We use the parametrization Eq.~\eqref{eq:bubbleMasterParam} to factor out the positive tadpole integral $I^d_{3,0}$, finally arriving at
\begin{align}
  & \frac {I^d_{3,0}} {M^2 (M^2 + 4m^2)} \big[ \left((6-d)M^2 + 4 m^2\right) \hat I^d_{2,1} - 4 m^2 \big] \geq 0 \, , \nonumber \\
  & \frac {I^d_{3,0}} {M^2 (M^2 + 4m^2)} \big[ -\left((8-d)M^2 + 12 m^2\right) \hat I^d_{2,1} + (2 M^2 + 12 m^2) \big] \geq 0 \, , \nonumber \\
  & \text{for any } d < 6 \, ,
\end{align}
i.e.
\begin{equation}
  \frac{4m^2}{(6-d)M^2 + 4 m^2} \leq \hat I^d_{2,1} = I^d_{2,1} / I^d_{3,0} \leq \frac{2 M^2 + 12 m^2}{(8-d)M^2 + 12 m^2} \, , \quad \text{for any } d < 6 \, . \label{eq:bubbleCrudeResult}
\end{equation}
It is easily shown that the LHS of the inequality above is always less than the RHS when $d<6$, so the normalized bubble integral $\hat I^d_{2,1} = I^d_{2,1} / I^d_{3,0}$ is bounded in a finite range. When $d$ tends to $6$ from below, the LHS and RHS of the inequality both approach $1$, leading to the prediction that $I^d_{2,1} / I^d_{3,0} \to 1$ as $d \to 6$; this is exactly as expected since both $I^d_{2,1}$ and $I^d_{3,0}$ have an ultraviolet pole $1/(d-6)$ with the same coefficient. Specializing to the case $d=4$, Eq.~\eqref{eq:bubbleCrudeResult} becomes
\begin{equation}
  \frac{2m^2}{M^2 + 2 m^2} \leq \hat I^{d=4}_{2,1} \leq \frac{M^2 + 6 m^2}{2M^2 + 6 m^2} \, , \quad \text{for } d = 4 \, . \label{eq:bubbleCrudeResult4D}
\end{equation}
Let us arbitrarily choose an example numerical point,
\begin{equation}
  M^2 = 2, \quad m^2 = 1 \, , \label{eq:bubbleEucNumPoint}
\end{equation}
At this point, Eq.~\eqref{eq:bubbleCrudeResult4D} becomes
\begin{equation}
  0.5 \leq \hat I^{d=4}_{2,1} \leq 0.8 \, , \quad \text{for } M^2=2, m^2=1 \, . \label{eq:bubbleCrudeResultNum}
\end{equation}
This is consistent with the analytic result given in Appendix \ref{app:bubbleAnalytic} with $p^2 = -M^2$,
\begin{align}
  \hat I^{d=4}_{2,1}  &= I^{d=4}_{2,1} / I^{d=4}_{3,0} = \frac {2 m^2} {\beta M^2} \log \frac {\beta+1}{\beta-1} \, , \label{eq:bubbleAnalyticFinite} \\
  \beta & \equiv \sqrt{ 1 + \frac {4m^2}{M^2}} \ ,
\end{align}
which evaluates to
\begin{equation}
  \hat I^{d=4}_{2,1} \approx 0.7603459963 \, , \quad \text{at } M^2=2, m^2=1 \, . \label{eq:bubEucAnalytic}
\end{equation}

\begin{figure}
  \centering
  \includegraphics[width=0.6\textwidth]{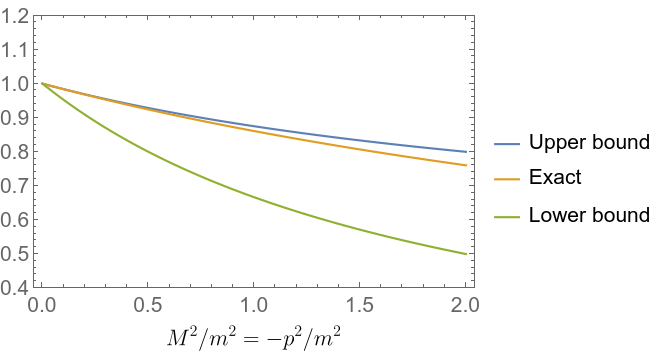}
  \caption{Comparison between ad hoc positivity bounds Eq.~\eqref{eq:bubbleCrudeResult4D} and the analytic result for the bubble integral $\hat I^{d=4}_{2,1}$ normalized according to Eq.~\eqref{eq:bubbleNormalizedA1A2}.}
  \label{fig:boundPlot}
\end{figure}

Since our crude positivity bound Eq.~\eqref{eq:bubbleCrudeResultNum} and the analytic result Eq.~\eqref{eq:bubbleAnalyticFinite} only depend on the dimensionless ratio $M^2/m^2$, we plot them in Fig.~\ref{fig:boundPlot}. It is not surprising that all three curves in the plot tend to $1$ as $M^2/m^2 \to 0$, since in this case we can set the external momenta to 0, and the bubble and tadpole integrals in Eq.~\eqref{eq:bubbleFiniteBasis} then become identical. The general observation is that our positivity bounds can become exact in special limits of kinematics or the spacetime dimension.
\subsubsection{Formulation in terms of matrix eigenvalues}
\label{subsubsec:eig}
In preparation for the introduction of semidefinite programming, we will first recast positivity constraints in terms of eigenvalues of an appropriate matrix. To simplify the notation of Eq.~\eqref{eq:selfEnergyInt}, let us define the following shorthand notations for the denominators,
\begin{equation}
  \rho_1 = -l^2 + m^2, \qquad \rho_2 = -(p+l)^2 + m^2 \, .
\end{equation}
Let us consider a class of positive-semidefinite integrals of the bubble family, parametrized by three real numbers $\alpha_1, \alpha_2, \alpha_3$,
\begin{equation}
  \int \frac {d^d \bm l \, e^{\gamma_E \epsilon}}{i \pi^{d/2}} \frac{1} {\rho_1^2 \rho_2}
  \left( \alpha_1 + \frac{\alpha_2}{\rho_1} + \frac{\alpha_3}{\rho_2} \right)^2
  = \sum\limits_{i,j} \alpha_i M_{ij} \alpha_j \\
  = {\vec \alpha}^T \, \mathbb M \, \vec \alpha \, , \label{eq:posAnsatzSize3}
\end{equation}
where the last line switched to a notation involving a length-3 column vector
\begin{equation}
  \vec \alpha = \begin{pmatrix} \alpha_1 \\ \alpha_2 \\ \alpha_3 \end{pmatrix}
\end{equation}
and a $3\times 3$ symmetric matrix $\mathbb M$, given by
\begin{equation}
  \mathbb M = \begin{pmatrix}
    I^d_{2,1} & I^d_{3,1} & I^d_{2,2} \\
    I^d_{3,1} & I^d_{4,1} & I^d_{3,2} \\
    I^d_{2,2} & I^d_{3,2} & I^d_{2,3}
  \end{pmatrix} \, , \label{eq:posMatrix3by3}
\end{equation}
using the index notation Eq.~\eqref{eq:selfEnergyInt}.
The expression Eq.~\eqref{eq:posAnsatzSize3} is non-negative for any choice of $(\alpha_1, \alpha_2, \alpha_3)$ because after Wick rotation, $\rho_1$ and $\rho_2$ are non-negative and the squared expression is also non-negative. Therefore, the symmetric matrix $\mathbb M$ must be positive-semidefinite, represented by the shorthand notation
\begin{equation}
  \mathbb M \succcurlyeq 0 \, , \label{eq:posGeq} \, .
\end{equation}

By IBP reduction, the matrix entries of $\mathbb M$ are integrals which can be re-expressed as linear sums of the two finite master integrals on the LHS of Eq.~\eqref{eq:bubbleFiniteBasis}. As an example,
\begin{align}
  \mathbb M_{23} &= \mathbb M_{32} = \int \frac {d^d \bm l \, e^{\gamma_E \epsilon}}{i \pi^{d/2}} \frac{1} {\rho_1^2 \rho_2^2} = I^d_{2,2},
\end{align}
which is then reduced to the finite master integrals according to Eq.~\eqref{eq:I22result}. Therefore $\mathbb M$ can be written as the sum of two individual master integral contributions,
\begin{equation}
  \mathbb M = I_{3,0}^d \mathbb M_1 + I_{2,1}^d \mathbb M_2 = I_{3,0}^d \left( \mathbb M_1 + \hat I^d_{2,1} \mathbb M_2 \right) \, , \label{eq:M1plusM2}
\end{equation}
using the notation $\hat I^d_{2,1} = I^d_{2,1} / I^d_{3,0}$ introduced in Eq.~\eqref{eq:bubbleMasterParam}. Since $I_{3,0}$ is itself positive, the positive-semidefiniteness of $\mathbb M$ implies
\begin{equation}
  \widetilde {\mathbb M} \equiv \mathbb M / I^d_{3,0} = \mathbb M_1 + \hat I^d_{2,1} \mathbb M_2 \succcurlyeq 0 \, , \label{eq:tildeMpos}
\end{equation}
again employing the shorthand notation introduced in Eq.~\eqref{eq:posGeq} to indicate positive-semidefiniteness. Equivalently, all the eigenvalues of $\mathbb M_1 + \hat I^d_{2,1} \mathbb M_2$ must be non-negative.
In Eq.~\eqref{eq:M1plusM2}, the matrices $\mathbb M_1$ and $\mathbb M_2$ contain entries that are rational functions of the spacetime $d$ and kinematic variables $p^2, m^2$, since IBP reduction always produces rational coefficients for master integrals. Although the general $d$ dependence is not complicated, we will present $\mathbb M_1$ and $\mathbb M_2$ in the $d=4$ case for brevity of presentation,
\begin{align}
  \mathbb M_1 \big|_{d=4} &=
  \begin{pmatrix}
    0 & \frac{2 m^2+M^2}{M^2 \left(4 m^2+M^2\right)} & -\frac{4m^2}{M^2 \left(4 m^2+M^2\right)} \\
    \frac{2 m^2+M^2}{M^2 \left(4 m^2+M^2\right)} & \frac{\left(m^2+M^2\right) \left(6 m^2+M^2\right)}{3 m^2 M^2 \left(4 m^2+M^2\right)^2} & \frac{M^2 - 2m^2}{M^2 \left(4 m^2+M^2\right)^2} \\
    -\frac{4 m^2}{M^2 \left(4 m^2+M^2\right)} & \frac{M^2-2 m^2}{M^2 \left(4 m^2+M^2\right)^2} & -\frac{2 m^2-M^2}{M^2 \left(4 m^2+M^2\right)^2} \\
  \end{pmatrix} \, , \\
  \mathbb M_2 \big|_{d=4}  &=
  \begin{pmatrix}
    1 & -\frac{2 m^2}{M^2 \left(4 m^2+M^2\right)} & \frac{2 \left(2 m^2+M^2\right)}{M^2 \left(4 m^2+M^2\right)} \\
    -\frac{2 m^2}{M^2 \left(4 m^2+M^2\right)} & -\frac{2 m^2}{M^2 \left(4 m^2+M^2\right)^2} & \frac{2 \left(m^2+M^2\right)}{M^2 \left(4 m^2+M^2\right)^2} \\
    \frac{2 \left(2 m^2+M^2\right)}{M^2 \left(4 m^2+M^2\right)} & \frac{2 \left(m^2+M^2\right)}{M^2 \left(4 m^2+M^2\right)^2} & \frac{2 \left(m^2+M^2\right)}{M^2 \left(4 m^2+M^2\right)^2} \\
  \end{pmatrix} \, .
\end{align}

\begin{figure}
  \centering
  \includegraphics[width=0.5\textwidth]{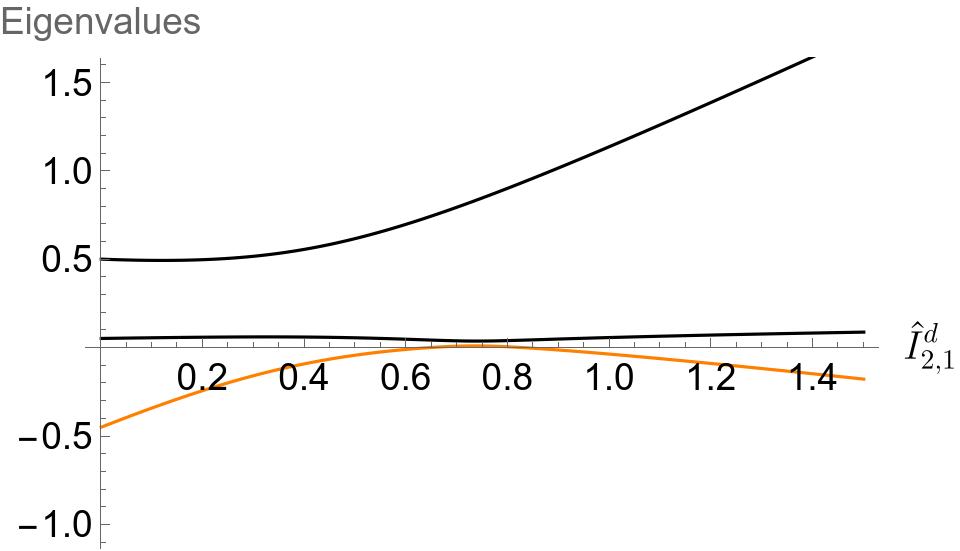}
  \caption{Three eigenvalues of $\widetilde {\mathbb M}$ defined in Eq.~\eqref{eq:tildeMpos} as a function of $\hat I^d_{2,1}$, at the kinematic point Eq.~\eqref{eq:bubbleEucNumPoint}. The lowest eigenvalue corresponds to the bottom orange curve and the remaining two eigenvalues correspond to the upper black curves.}
  \label{fig:momSpaceThreeEigs}
\end{figure}

\begin{figure}
  \centering
  \includegraphics[width=0.5\textwidth]{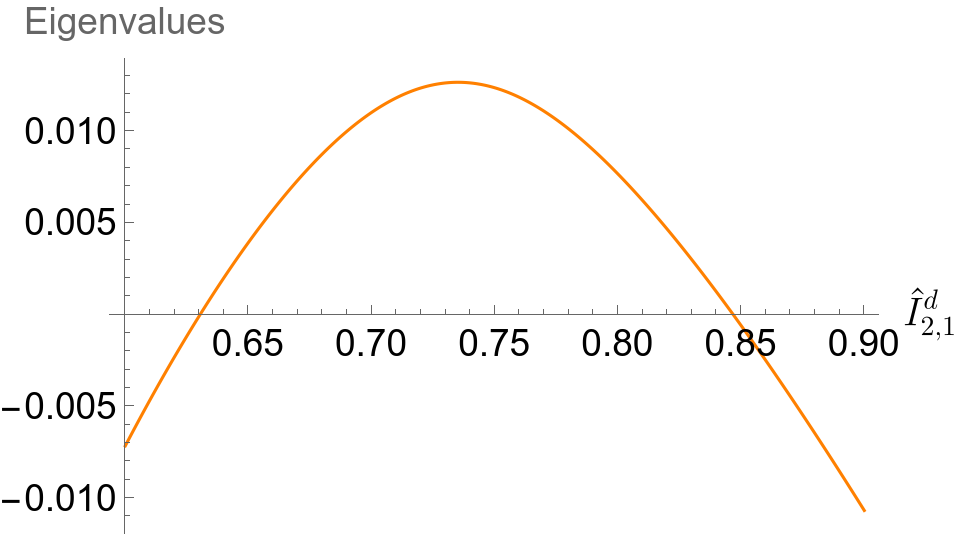}
  \caption{Magnified version of the vicinity of a small region of Fig.~\ref{fig:momSpaceThreeEigs} in which the lowest eigenvalue of $\widetilde {\mathbb M}$, shown in the curve, is non-negative.}
  \label{fig:momSpaceThreeEigs1}
\end{figure}

\begin{figure}
  \centering
  \includegraphics[width=0.5\textwidth]{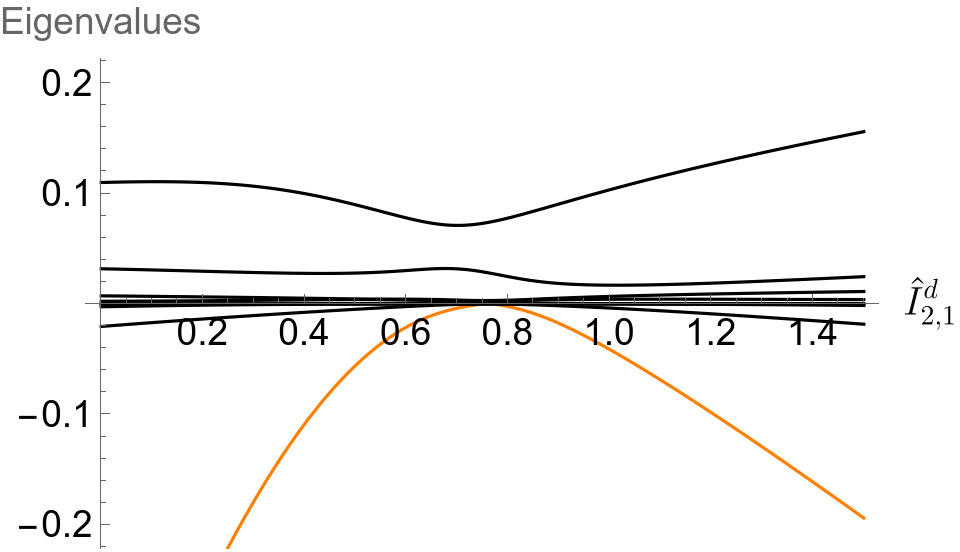}
  \caption{Eigenvalues as a function $\hat I^d_{2,1}$, for the $10\times 10$ symmetric matrix that represent the quadratic dependence of the RHS of Eq.~\eqref{eq:posAnsatzSize10} on the $\alpha_i$ parameters after factoring out $I^d_{3,0}$.}
  \label{fig:momSpaceTenEigs}
\end{figure}

\begin{figure}
  \centering
  \includegraphics[width=0.5\textwidth]{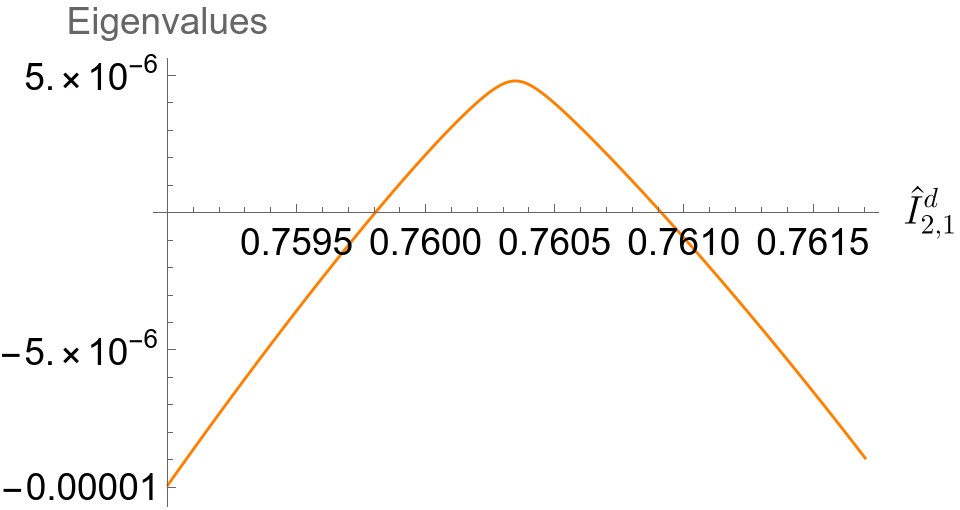}
  \caption{A magnified version of Fig.~\ref{fig:momSpaceTenEigs} around the small region of $\hat I^d_{2,1}$ where all eigenvalues are positive, showing only the smallest eigenvalue.}
  \label{fig:momSpaceTenEigs1}
\end{figure}

Now we treat $\hat I^d_{2,1}$ as an undetermined parameter to be constrained by Eq.~\eqref{eq:tildeMpos}. Again we look at the example numerical point as in Eq.~\eqref{eq:bubbleEucNumPoint},
\begin{equation}
  M^2 = 2, \quad m^2 = 1 \, ,
\end{equation}
and plot the three eigenvalues of the $3\times 3$ matrix $\widetilde {\mathbb M} = \mathbb M_1 + \hat I^d_{2,1} \mathbb M_2$ in Fig.~\ref{fig:momSpaceThreeEigs}.
It can be seen in the figure that most of the parameter range shown is ruled out due to the presence of a negative eigenvalue indicated by the lowest orange curve. In Fig.~\ref{fig:momSpaceThreeEigs1}, we zoom in to a smaller parameter region and only plot the smallest eigenvalue, since the matrix is positive semidefinite as long as the smallest eigenvalue is non-negative.
The allowed parameter region shown in the plot is
\begin{equation}
  0.630 \leq \hat I^{d=4}_{2,1} \leq 0.847 \, ,
\end{equation}
which provides a more stringent lower bound than the previous result Eq.~\eqref{eq:bubbleCrudeResultNum}. In fact, far better lower and upper bounds can be achieved in this approach by using an ansatz larger than the one in Eq.~\eqref{eq:posAnsatzSize3}. For example, let us replace the squared term in Eq.~\eqref{eq:posAnsatzSize3} by an arbitrary degree-3 polynomial in $1/\rho_1$ and $1/\rho_2$, parametrized by 10 undetermined free coefficients. We obtain the constraint,
\begin{equation}
  0 <\int \frac {d^d \bm l \, e^{\gamma_E \epsilon}}{i \pi^{d/2}} \frac{1} {\rho_1^2 \rho_2}
  \left(\alpha_1 + \frac{\alpha_2}{\rho_1} + \frac{\alpha_3}{\rho_2}
  + \frac{\alpha_4}{\rho_1^2} + \frac{\alpha_5}{\rho_1 \rho_2} + \frac{\alpha_6}{\rho_2^2}
  + \frac{\alpha_7}{\rho_1^3} + \frac{\alpha_8}{\rho_1^2 \rho_2} + \frac{\alpha_9}{\rho_1 \rho_2^2} + \frac{\alpha_{10}}{\rho_2^3}
  \right)^2 \, . \label{eq:posAnsatzSize10}
\end{equation}
Repeating the above analysis, we obtain a constraint for $\hat I^d_{2,1}$ similar to Eq.~\eqref{eq:tildeMpos}, except that the matrices involved have $10 \times 10$ sizes. The ten eigenvalues as functions of $\hat I^d_{2,1}$ are plotted in Fig.~\ref{fig:momSpaceTenEigs}, as an ``upgraded'' version of Fig.~\ref{fig:momSpaceThreeEigs} with a larger ansatz size. Some of the ten eigenvalues are too close to each other on the plot to be seen individually, but only the smallest eigenvalue (represented by the lowest curve in orange color) matters as it determines whether we can satisfy the constraint that all eigenvalues are positive.
In Fig.~\eqref{fig:momSpaceTenEigs1}, we zoom in to the small allowed range for the parameter $\hat I^d_{2,1}$.
The allowed region, as shown in the plot, is
\begin{equation}
  0.7598 \leq \hat I^{d=4}_{2,1} \leq 0.7610 \, ,
\end{equation}
which tightly constrains $\hat I^{d=4}_{2,1}$ around its true value $\hat I^{d=4}_{2,1} \approx 0.7603$ from evaluating the known analytic result at $d=4, M^2 = -p^2 = 2$, with a relative error of around $10^{-3}$.

The above two-sided bounds are rigorous, but we will also explore a prescription to assign a ``central value'', or ``best estimate'', of the value of the integrals. The prescription described below, though not justified from first principles, empirically achieve a closer agreement with true values of the integrals than the rigorous bounds in the examples in this paper. The prescription is simply finding the value of $\hat I^d_{2,1}$ which maximizes the smallest eigenvalue of the matrix that is required to be positive semidefinite, e.g.\ the matrix $\widetilde M$ of Eq.~\eqref{eq:M1plusM2}. For the positivity constraint Eq.~\eqref{eq:posAnsatzSize10} with 10 free parameters, the prescription picks the value of $\hat I^d_{2,1}$ corresponding to the maximum of the curve in Fig.~\ref{fig:momSpaceTenEigs1}, which deviates from the exact result, again at the example point $d=4, m=1, M^2=-p^2=2$, by only a relative error of about $10^{-6}$. In this case, the prescription happens to produce a value that is very close to the exact result, but typically we observe the prescription to give a ``central value'' that is one to two orders of magnitude better than the accuracy indicated by rigorous bounds.

\subsubsection{High-precision evaluation using semidefinite programming}
\label{subsubsec:sdp}
We have formulated positivity constraints in terms of eigenvalues of a matrix which, in the toy example of the one-loop bubble integrals, depends linearly on only one undetermined parameter, as shown in Eq.~\eqref{eq:tildeMpos}. For more complicated Feynman integrals, there will be more than one master integrals to be evaluated and all of them will be considered as undetermined parameters. So a search in higher-dimensional space is needed to locate the region in which all eigenvalues of the matrix are non-negative, and this can become computationally challenging. Fortunately, very efficient algorithms exist to solve \emph{semidefinite programming problems} in mathematical optimization \cite{vandenberghe1996semidefinite}. Loosely speaking, semidefinite programs are generalizations of linear programs allowing not only linear constraints but also positive semidefiniteness constraints on matrices that have linear dependence on the optimization variables. Here we show how our problem of constraining unknown master integrals can be stated as semidefinite programming problems. Following the treatment of Section \ref{subsubsec:eig} above, finding the minimum allowed value of $\hat I^d_{2,1}$ can be formulated as
\begin{equation}
\begin{aligned}
\text{minimize } \quad & \hat I^d_{2,1} \, , \\
\text{subject to }\quad & \mathbb M_1 + \hat I^d_{2,1} \cdot \mathbb M_2 \succcurlyeq 0 \, ,
\end{aligned}
\label{eq:sdpMin}
\end{equation}
which is in the form of a semidefinite program. We used the $\succcurlyeq 0$ notation, already introduced in Eq.~\eqref{eq:posGeq}, to indicate that the matrix on the LHS must be positive semidefinite.
Similarly, finding the maximum allowed value of $\hat I^d_{2,1}$ can be formulated as
\begin{equation}
\begin{aligned}
\text{maximize } \quad & \hat I^d_{2,1} \, , \\
\text{subject to }\quad & \mathbb M_1 + \hat I^d_{2,1} \cdot \mathbb M_2 \succcurlyeq 0 \, .
\end{aligned}
\label{eq:sdpMax}
\end{equation}
Finally, to implement our prescription of maximizing the smallest eigenvalue to find the ``central value'' of the undetermined master integrals, we introduce an additional undetermined parameter $\lambda$ and formulate the problem as
\begin{equation}
\begin{aligned}
\text{maximize } \quad & \lambda \, , \\
\text{subject to }\quad & \mathbb M_1 + \hat I^d_{2,1}\cdot \mathbb M_2 - \lambda \mathbb I \succcurlyeq 0 \, .
\end{aligned}
\label{eq:sdpCentral}
\end{equation}
This is again in the form of a semidefinite program, where both $\hat I^d_{2,1}$ and $\lambda$ are undetermined parameters whose values will be fixed to satisfy the optimization objective, namely to maximize $\lambda$. Note that in this case, finding a optimal solution to the semidefinite program does not guarantee $\mathbb M_1 + \hat I^d_{2,1} \cdot \mathbb M_2 \succcurlyeq 0$, \emph{unless} the value of $\lambda$ in the solution is non-negative. The value of $\hat I^d_{2,1}$ in the solution is then taken as the central value for this undetermined free parameter.

There exist many computer codes that specialize in solving semidefinite programs. Wolfram Mathematica has supported semidefinite programming since version 12 with the {\tt SemidefiniteOptimization} function, and the default backend (which can be changed by the {\tt Method} option) is the open source library CSDP \cite{borchers1999csdp} at least for the problems we deal with, working with double-precision floating numbers, i.e.\ the standard machine precision on current hardware. The SDPA family \cite{yamashita2012latest} of computer programs support computation at double precision as well as a variety of extended precisions, e.g.\ double-double precision with SDPA-DD, quadruple-double precision with SDPA-QD, and arbitrary precision with SDPA-GMP. The SDPB solver by Simmons-Duffin \cite{Simmons-Duffin:2015qma} specializes in polynomial programming problems in the conformal bootstrap and works in arbitrary precision. Most of the work in this paper will make use of the SDPA family, while some results from Mathematica / CSDP will also be shown for the purpose of comparison.

\begin{figure}
  \centering
  \includegraphics[width=0.8\textwidth]{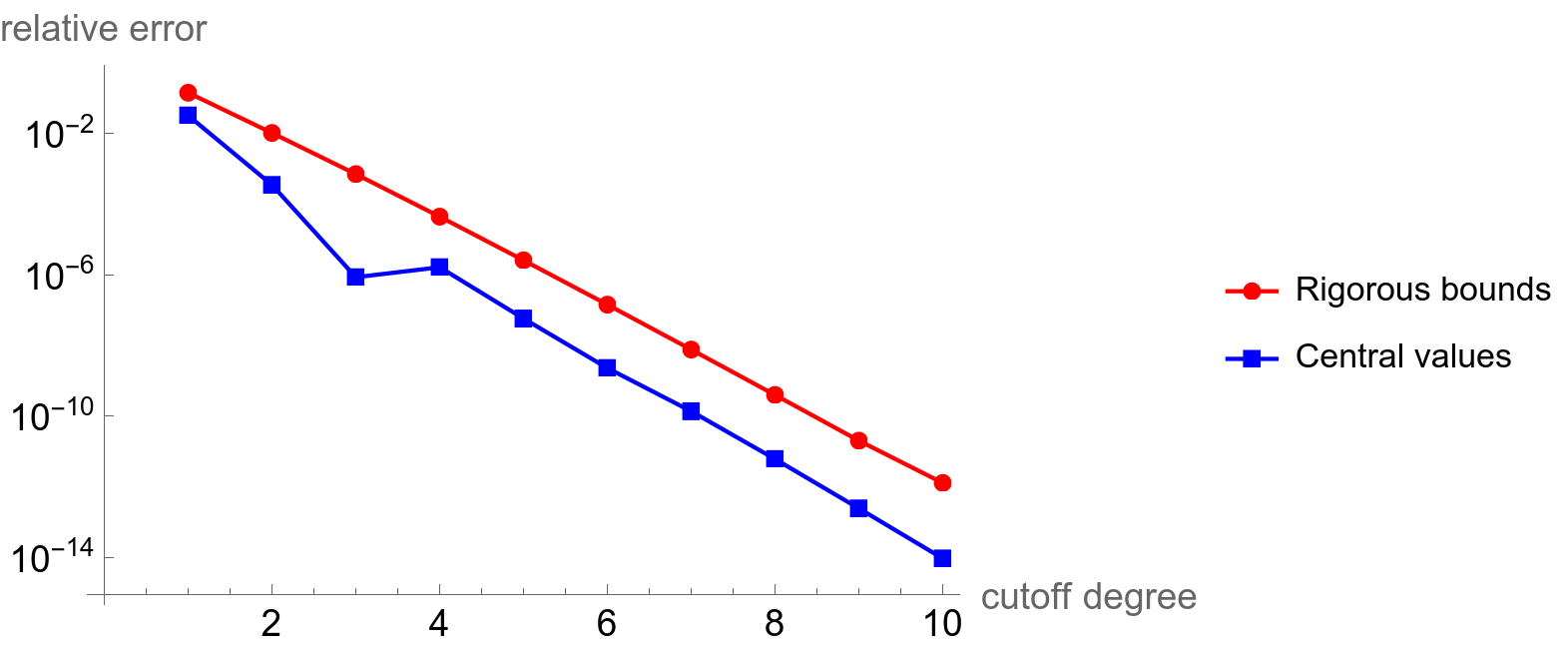}
  \caption{Relative errors in numerical results for $\hat I^{d=4}_{2,1}$ at the kinematic point Eq.~\eqref{eq:bubbleEucNumPoint} from solving positivity constraints Eq.~\eqref{eq:posAnsatzSizeGeneral} using semidefinite programming. The horizontal axis is the cutoff degree of the polynomial $P$. The relative errors are defined by Eq.~\eqref{eq:defErrorRigorous} for the rigorous bounds and Eq.~\eqref{eq:defErrorCentral} for the central values.}
  \label{fig:error_bounds_central}
\end{figure}

We go on to discuss how to achieve higher numerical precision for the one-loop bubble integral. We enlarge the ansatz for positive integrals in Eqs.~\eqref{eq:posAnsatzSize3} and \eqref{eq:posAnsatzSize10} to have more parameters, as
\begin{equation}
  0 < \int \frac {d^d \bm l \, e^{\gamma_E \epsilon}}{i \pi^{d/2}} \frac{1} {\rho_1^2 \rho_2}
  P(1/\rho_1, 1/\rho_2)^2 \, , \label{eq:posAnsatzSizeGeneral}
\end{equation}
where P is an arbitrary polynomial with a maximum degree $N$, i.e.\ an arbitrary linear sum of all monomials in $1/\rho_1$ and $1/\rho_2$, with each monomial multiplied by a free parameter $\alpha_i$. The $N=1$ and $N=3$ cases are shown previously in Eqs.~\eqref{eq:posAnsatzSize3} and \eqref{eq:posAnsatzSize10}, respectively.  Generally, the number of possible monomials in two variables with maximum degree $N$ is equal to $(N+1)(N+2)/2$. For each value of $N$ from 1 to 10, we again set $\hat I^d_{2,1} = I_{2,1} / I_{3,0}$ at $d=4, m=1, M^2=-p^2=2$ and solve the semidefinite programs in Eqs.~\eqref{eq:sdpMin} to \eqref{eq:sdpCentral} to obtain the lower bound $(\hat I^d_{2,1})_{\rm min}$, upper bound $(\hat I^d_{2,1})_{\rm max}$, and central value $(\hat I^d_{2,1})_{\rm central}$ for the undetermined parameter $\hat I^d_{2,1}$. To ensure numerical stability, we use the SDPA-DD solver working at double-double precision. Then we compare with the exact result $(\hat I^d_{2,1})_{\rm exact}$ to find the relative error in the best estimate value, defined as
\begin{equation}
  \left| \frac {(\hat I^d_{2,1})_{\rm central}} {(\hat I^d_{2,1})_{\rm exact}} - 1 \right|
  \label{eq:defErrorCentral}
\end{equation}
as well as the relative error of the rigorous bounds, defined as
\begin{equation}
  \left| \frac{ (\hat I^d_{2,1})_{\rm max} - (\hat I^d_{2,1})_{\rm min} }{ 2 (\hat I^d_{2,1})_{\rm exact}} \right|
  \label{eq:defErrorRigorous}
\end{equation}
In Fig.~\ref{fig:error_bounds_central}, we plot the relative errors against the cutoff degree $N$ of the polynomial $P$ in Eq.~\eqref{eq:posAnsatzSizeGeneral}.
The plot is on a log scale, and we can see that the numerical results appear to converge exponentially to the true value, reaching a precision as high as $10^{-14}$ with a cutoff degree of 10. The central value from our somewhat arbitrary prescription is seen to be consistently more precise than the rigorous bounds.

\begin{figure}
  \centering
  \includegraphics[width=0.8\textwidth]{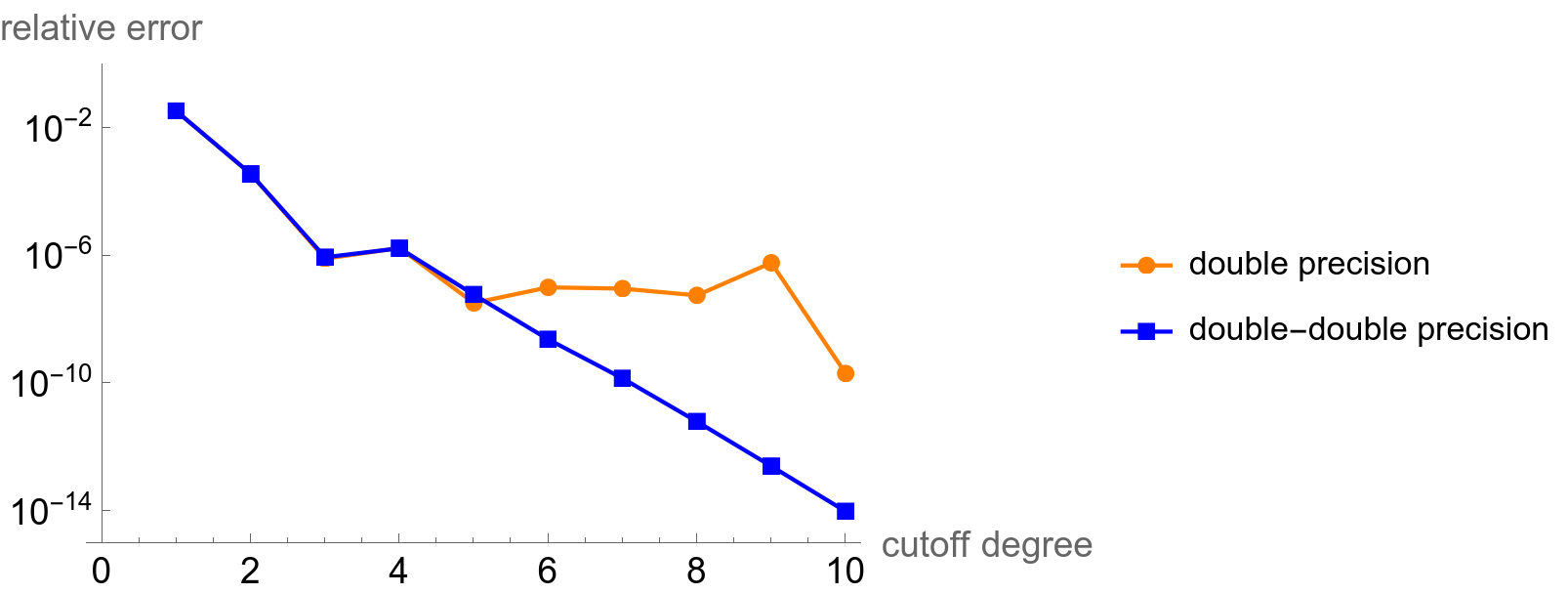}
  \caption{Relative errors in numerical results for the central values of $\hat I^{d=4}_{2,1}$ at the kinematic point Eq.~\eqref{eq:bubbleEucNumPoint}, obtained with semidefinite programming solvers working at two different numerical precisions, namely Mathematica / CSDP working at double precision and SDPA-DD working at double-double precision. As the plot shows, double-double precision is needed for the relative errors to improve exponentially beyond a cutoff degree of 5.}
  \label{fig:error_CSDP_SDPA_DD}
\end{figure}

We revisit the issue of numerical stability. In Fig.~\ref{fig:error_CSDP_SDPA_DD}, we compare the accuracy obtained for the central values obtained by SDPA-DD with double-double precision, which was used above, and Mathematica / CSDP with double precision. The two results visibly deviate from each other once the cutoff degree is 5 or above, and only the double-double precision computation continues to exhibit exponential reduction in errors as the cutoff degree is increased. This signals that numerical instability has occurred if one computes at double precision only. We have checked that the accuracies do not improve further when computing with quadruple-double precision using SDPA-QD.

\subsection{Positivity constraints in Feynman parameter space}
\label{subsec:posFeynSpace}
In Section \ref{subsec:posMomSpace}, we used positivity constraints in loop momentum space to evaluate bubble integrals defined in Eq.~\eqref{eq:selfEnergyInt} in the case $p^2 < 0$ when it is possible to Wick-rotate the integrals into Euclidean spacetime with real-valued external momenta. We now use positivity constrains in Feynman parameter space instead to evaluate bubble integrals for any value of $p^2$ less than $4m^2$, which is what is commonly referred to as the ``Euclidean region'' where the bubble integrals have no imaginary parts.
We write down the Feynman parameter representation of bubble integrals defined in Eq.~\eqref{eq:selfEnergyInt},
\begin{align}
  I_{a_1,a_2}^d & \equiv \int \frac {d^d l \, e^{\gamma_E \epsilon}}{i \pi^{d/2}} \frac{1} {(- l^2 + m^2)^{a_1} [-(p + l)^2 + m^2]^{a_2}} \nonumber \\
  &= \frac {\Gamma(a_1 + a_2 - d / 2) e^{\gamma_E \epsilon}}
  {\Gamma(a_1) \Gamma(a_2)} \int_0^\infty dx_1 \int_0^\infty dx_2 \, \delta (1-x_1-x_2) \nonumber \\
  & \quad \times x_1^{a_1 - 1} x_2^{a_2 - 1}
  \frac {\mathcal U(x_1, x_2)^{a_1 + a_2 -d}} {\mathcal F(x_1, x_2)^{a_1 + a_2 - d/2}} \label{eq:bubbleFeynParam0} \\
  &= \frac {\Gamma(a_1 + a_2 - d / 2) e^{\gamma_E \epsilon}}
  {\Gamma(a_1) \Gamma(a_2)} \int_0^1 dx \, x^{a_1 - 1} (1-x)^{a_2 - 1}
  \frac 1 {\mathcal F(x)^{a_1 + a_2 - d/2}} \, ,
  \label{eq:bubbleFeynParam}
\end{align}
where the graph polynomials were already given in Eq.~\eqref{eq:bubbleFpoly0} for the Schwinger parametrization, printed again here:
\begin{equation}
  \mathcal U(x_1, x_2) = x_1+x_2, \quad \mathcal F(x_1, x_2) = m^2(x_1+x_2)^2 - p^2 x_1 x_2 - i 0^+ \, .
  \label{eq:bubbleFpoly0}
\end{equation}
In the last line of Eq.~\eqref{eq:bubbleFeynParam}, we integrated $x_2$ over the delta function in Eq.~\eqref{eq:bubbleFeynParam0} to arrive at Eq.~\eqref{eq:bubbleFeynParam} with
\begin{equation}
  \mathcal F(x) \equiv \mathcal F(x, 1-x) = m^2 - p^2 x(1-x) - i 0^+ \, ,
  \label{eq:bubbleFpoly}
\end{equation}
and
\begin{equation}
  \mathcal U(x) \equiv \mathcal U(x, 1-x) \equiv 1 \, ,
\end{equation}
which appears as a unit numerator in Eq.~\eqref{eq:bubbleFeynParam}.

\emph{Aside:} we note that Eq.~\eqref{eq:bubbleFeynParam0} has the property that when ignoring the Dirac delta function $\delta(1-\sum_i x_i)$, the rest of the expression (with the integration measure taken into account) is invariant under the rescaling
\begin{equation}
  x_i \to \lambda x_i,
  \label{eq:projective}
\end{equation}
where $\lambda$ is an arbitrary nonzero real number. This is called \emph{projective invariance} and holds for the Feynman parameter form of arbitrary Feynman integrals written down in Eq.~\eqref{eq:generalFeynParam}. The deeper reason is that Feynman parameter integrals can generally be written as integrals in real projective space $\mathbb R \mathbb P^{N-1}$ (see e.g.\ Section 2.5.3 of Ref.~\cite{Weinzierl:2022eaz}). $\mathbb R \mathbb P^{N-1}$ is the space of $N$ real coordinates $x_i$, excluding the origin, where any ray, i.e.\ a set of points related to each other by a rescaling Eq.~\eqref{eq:projective}, is identified as the same point.
Abusing the language of gauge theory, Eq.~\eqref{eq:projective} is a gauge symmetry and $\delta(1- \sum_i x_i)$ in Eq.~\eqref{eq:bubbleFeynParam0} is a gauge-fixing term that restricts the integration to one of the infinitely many gauge-equivalent slices. The Fadeev-Popov Jacobian associated with this gauge-fixing term is unity and therefore does not appear explicitly. Projective invariance is not a prerequisite for following the rest of the paper, though it helps motivate some of the developments.

Now we specialize to the following kinematic region for bubble integrals,
\begin{equation}
  0 < p^2 < 4m^2,
\end{equation}
i.e. with the value of $p^2$ below the Cutkosky cut threshold but cannot be trivially Wick-rotated into Euclidean spacetime. We have
\begin{equation}
  m^2 - p^2 x (1-x) \geq m^2 - p^2/4 > 0,
\end{equation}
so the $-i0^+$ prescription in Eq.~\eqref{eq:bubbleFpoly} is negligible and can be dropped, and the integral is real.
For the rest of the paper, we will adopt the common terminology of the Euclidean region to be the kinematic region in which all graph polynomials are non-negative and the Feynman integral is real-valued due to the lack of Cutkosky cuts. Generally such integrals cannot be embedded into Euclidean spacetime. This is e.g.\ the working definition when the literature refers to the Euclidean region of Feynman integrals with massless external legs, since nonzero massless momenta cannot be literally embedded into Euclidean spacetime.

If we set $a_2 = 1$, we can invert Eq.~\eqref{eq:bubbleFeynParam} to obtain
\begin{align}
  &\quad \int_0^1 dx \, x^{a_1-1} \frac 1 {\mathcal F(x)^{1 + a_1 - d/2}} \nonumber \\
  &=
  \int_0^1 dx \, x^{a_1-1} \frac 1 {\left[ m^2 - p^2 x (1-x) \right]^{1 + a_1 - d/2}} \nonumber \\
  &= \frac{\Gamma(a_1)} {e^{\gamma_E \epsilon} \, \Gamma(a_1 + 1  - d/2)} I^d_{a_1, 1} \, ,
  \label{eq:bubbleFeynParamInvert1}
\end{align}

It will be more useful to have a version of the above equation with a fixed exponent for $\mathcal F(x)$ on the LHS, even when the value of $a_1$ changes. Below is a version with a fixed exponent $d/2 - 3$ for $\mathcal F(x)$, obtained by replacing $d \rightarrow d + 2(a_1-2)$ in Eq.~\eqref{eq:bubbleFeynParamInvert1} and multiplying by a constant prefactor,
\begin{align}
  & \quad 2(m^2)^{3-d/2} \int_0^1 dx \, x^{a_1-1} \frac 1 {\mathcal F(x)^{3-d/2}} \nonumber \\
  &=
  2 \int_0^1 dx \, x^{a_1-1} \frac 1 {\left[ 1 - p^2 x (1-x) / m^2 \right]^{3-d/2}} \nonumber \\
  &= \frac {2 \Gamma(a_1) (m^2)^{1+\epsilon}} {e^{\gamma_E \epsilon} \, \Gamma(3  - d/2)} I^{d+2a_1-4}_{a_1, 1} \nonumber \\
  &= I^{d+2a_1-4}_{a_1, 1} / I^d_{3, 0}
  \, , \label{eq:bubbleFeynParamInvert0}
\end{align}
where the last line used the explicit result for $I^d_{3,0}$ in Eq.~\eqref{eq:tadpoleAnalytic}. We define
\begin{equation}
  \hat F(x) = \mathcal F(x) / m^2 = 1 - p^2 x (1-x) / m^2 \, , \label{eq:bubbleHatFDef}
\end{equation}
and rewrite Eq.~\eqref{eq:bubbleFeynParamInvert0} as
\begin{equation}
  2 \int_0^1 dx \, x^{a_1-1} \frac 1 {\hat F(x)^{3-d/2}} = I^{d+2a_1-4}_{a_1, 1} / I^d_{3, 0} \, .
  \label{eq:bubbleFeynParamInvert}
\end{equation}
The RHS of Eq.~\eqref{eq:bubbleFeynParamInvert} can be simplified further, as IBP identities and dimension-shifting identities can be applied to reduce $I^{d+2a_1-4}_{a_1, 1}$ to a linear combination of the two master integrals, $I^d_{2,1}$ and $I^d_{3,0}$.
We will only use Eq.~\eqref{eq:bubbleFeynParamInvert} in the case $a_1 \geq 2$, when the RHS involves a bubble integral in spacetime dimension greater than or equal to $d= 4 - 2\epsilon$. As an example, consider the case $a_1=3$, and Eq.~\eqref{eq:bubbleFeynParamInvert} becomes
\begin{equation}
  2 \int_0^1 dx \, x^2 \frac 1 {\hat F(x)^{3-d/2}} = I^{d+2}_{3,1} / I^d_{3, 0} \, .
\end{equation}
Then we simplify the expression using the IBP reduction result Eq.~\eqref{eq:I31result} with $d$ replaced by $d+2$, obtaining
\begin{equation}
  2 \int_0^1 dx \, x^2 \frac 1 {\hat F(x)^{3-d/2}} = \frac{1}{2p^2 (4m^2-p^2)} \Big( [(2-d)p^2 + 4m^2] I^{d+2}_{2,1} + 2(p^2-2m^2) I^{d+2}_{3,0} \Big) / I^d_{3,0} \, .
\end{equation}
Finally, applying dimension-shifting identities Eqs.~\eqref{eq:dimshift1prime} and \eqref{eq:dimshift2prime}, the above equation becomes
\begin{align}
  2 \int_0^1 dx \, x^2 \frac 1 {\hat F(x)^{3 - d/2}} &= \left( \frac {(d-2)p^2 - 4m^2} {4(d-3)p^2} I^d_{2,1} + \frac{m^2}{(d-3) p^2} I^d_{3,0} \right) / I^d_{3,0} \nonumber \\
  &= \frac {(d-2)p^2 - 4m^2} {4(d-3)p^2} \hat I^d_{2,1} + \frac{m^2}{(d-3) p^2} \, .
  \label{eq:bubbleFeynParamInvertExample}
\end{align}
This concludes our example for simplifying the RHS of Eq.~\eqref{eq:bubbleFeynParamInvert} in the case of $a_1=3$.
We now formulate a first version of positivity constraints for any $d<6$, to be improved upon later, as,
\begin{equation}
  0 \leq 2 \int_0^1 dx \, x P(x)^2 \frac 1 {\hat F(x)^{3 - d/2}} \, ,
  \label{eq:posAnsatzFeynParam}
\end{equation}
where $P(x)$ is an arbitrary polynomial in $x$, which is analogous to the arbitrary polynomial $P(1/\rho_1, 1/\rho_2)$ in Eq.~\eqref{eq:posAnsatzSizeGeneral} used to construct positive integrals in momentum space. In the special case $P(x)=1$, the RHS of Eq.~\eqref{eq:posAnsatzFeynParam} is simply proportional to $I^d_{2,1}$ in unshifted spacetime dimension $d$, according to Eq.~\eqref{eq:bubbleFeynParamInvert}. Since $x$ is non-negative in the range of integration $0 \leq x \leq 1$, $x P(x)^2$ is non-negative.
We have chosen to use $x P(x)^2$ instead of just $P(x)^2$ to ensure that each monomial in the expanded expression contains at least one power of $x$ and is related to an ultraviolet convergent integral in Eq.~\eqref{eq:bubbleFeynParamInvert} as discussed above. Another valid choice is $(1-x) P(x)^2$, but this will not give more constraints for the bubble integral, because in Eq.~\eqref{eq:bubbleFeynParam}, $\mathcal F(x) = p^2 - m^2 x(1-x)$ is invariant under the exchange $x \leftrightarrow 1-x$, owing to a reflection symmetry of the bubble diagram.

It is possible to slightly refine the inequality Eq.~\eqref{eq:posAnsatzFeynParam} to make the constraint stronger. As we assume $p^2>0$, we have
\begin{equation}
  \mathcal F(x) = m^2 - p^2 x(1-x) < m^2, \quad
  \hat F(x) = \mathcal F(x) / m^2 < 1
  \, .
\end{equation}
So for any $d<6$, we can modify Eq.~\eqref{eq:posAnsatzFeynParam} with an extra term,
\begin{equation}
  0 \leq 2 \int_0^1 dx \, x P(x)^2 \left( \frac 1 {\hat F(x)^{3-d/2}} - 1 \right) \, . \label{eq:bubbleFeynRefinedPositivity}
\end{equation}
Eq.~\eqref{eq:bubbleFeynRefinedPositivity} can also be written in a form that manifests the projective invariance discussed around Eq.~\eqref{eq:projective},
\begin{align}
  0 &\leq 2 \int_0^\infty dx_1 \int_0^\infty dx_2 \, \delta(1-x_1-x_2) \nonumber \\
  &\quad \times \frac {x_1}{U(x_1, x_2)}  P\left( \frac {x_1}{\mathcal U(x_1, x_2)} \right)^2 \left( \frac {\mathcal U(x_1, x_2)^{3-d}} {\hat F(x)^{3-d/2}} - \frac 1 {\mathcal U(x_1, x_2)^3} \right) \, ,
\end{align}
but we will use the form Eq.~\eqref{eq:bubbleFeynRefinedPositivity} below.

To be more concrete, in Eq.~\eqref{eq:bubbleFeynRefinedPositivity}, we use
\begin{equation}
  P(x) = \alpha_1 + \alpha_2 x^2 + \dots + \alpha_N x^N \, , \label{eq:PxParametrization}
\end{equation}
where $N$ is the cutoff degree of the polynomial and $\alpha_i$ with $1 \leq i \leq N$ are free parameters, and Eq.~\eqref{eq:bubbleFeynRefinedPositivity} must hold for any values of the $\alpha_i$ parameters. For each monomial from expanding $x P(x)^2$, the first term in the curly bracket of Eq.~\eqref{eq:bubbleFeynRefinedPositivity} gives a bubble integral in a shifted dimension, normalized against the tadpole integral $I^d_{3,0}$, according to the formula Eq.~\eqref{eq:bubbleFeynParamInvert}, while the second term in the curly bracket of Eq.~\eqref{eq:bubbleFeynRefinedPositivity} contributes to an integral of a monomial in $x$ over $0\leq x \leq 1$ which can be evaluated trivially. Using both dimension shifting identities and IBP identities, the bubble integrals produced above are rewritten as linear combinations of finite master integrals Eq.~\eqref{eq:bubbleFiniteBasis}. Therefore Eq.~\eqref{eq:bubbleFeynRefinedPositivity} is turned into the form
\begin{equation}
  {\vec \alpha}^T \, \mathbb M \, \vec \alpha \geq 0 \, , \label{eq:aMa_is_positive}
\end{equation}
similar to the momentum-space version Eq.~\eqref{eq:posAnsatzSize3}, with
\begin{equation}
  \mathbb M = \mathbb M_1 + \hat I^d_{2,1} \mathbb M_2 \, ,
\end{equation}
where we use the definition $\hat I^d_{2,1} = I^d_{2,1} / I^d_{3,0}$ as before, and the ``inhomogeneous'' term $\mathbb M_1$ receives contribution from both constant terms in Eq.~\eqref{eq:bubbleFeynRefinedPositivity} and tadpole integrals coming from dimension-shifting and IBP identities. Following analogous developments in Sections \ref{subsubsec:eig} and \ref{subsubsec:sdp}, we solve the constraint
\begin{equation}
  \mathbb M_1 + \hat I^d_{2,1} \mathbb M_2 \succcurlyeq 0
\end{equation}
while minimizing or maximizing $\hat I^d_{2,1}$ to find rigorous bounds for $\hat I^d_{2,1}$, or alternatively maximizing the smallest eigenvalue of $\mathbb M_1 + \hat I^d_{2,1} \mathbb M_2$ to find a central value for $\hat I^d_{2,1}$ using the same prescription as described before.

\begin{figure}
  \centering
  \includegraphics[width=0.8\textwidth]{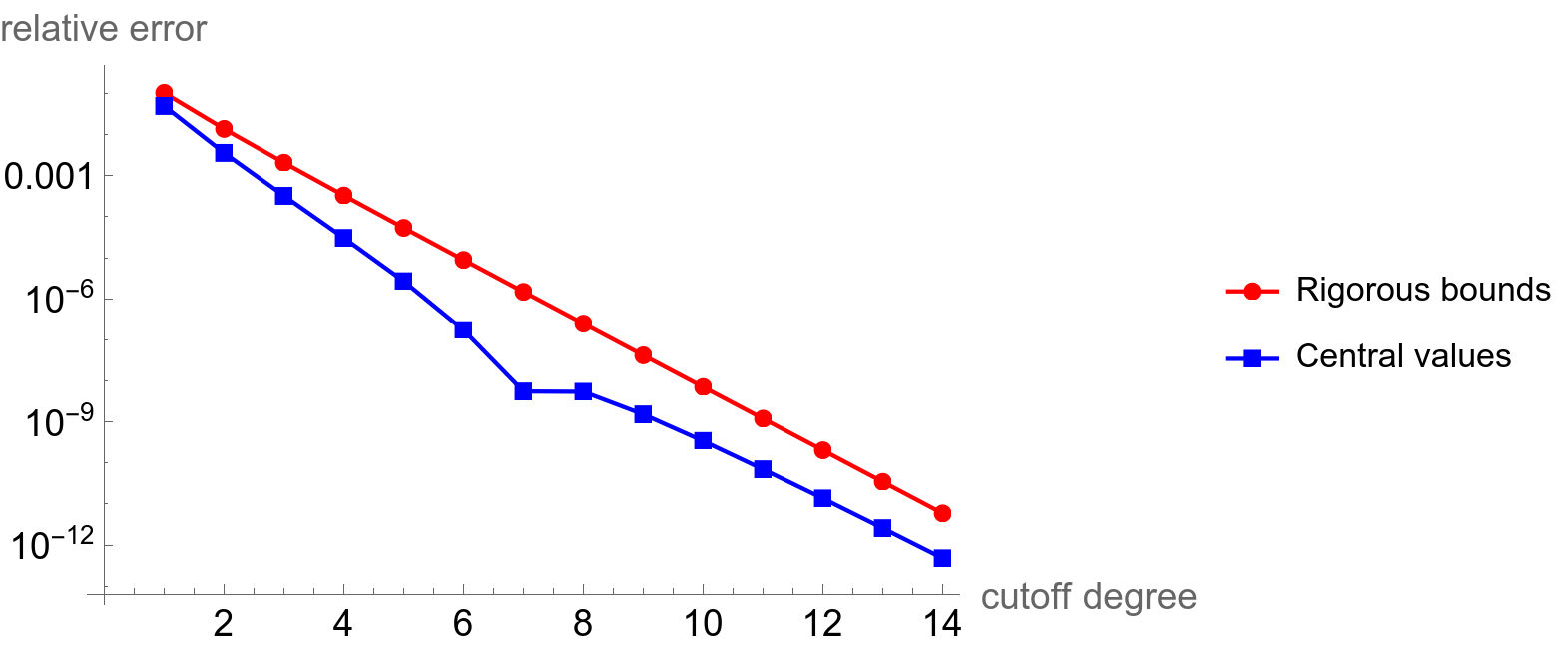}
  \caption{Relative errors in numerical results for $\hat I^{d=4}_{2,1}$ at the kinematic point Eq.~\eqref{eq:bubbleNumPoint} from solving positivity constraints Eq.~\eqref{eq:bubbleFeynRefinedPositivity} in Feynman-parameter space using semidefinite programming. The horizontal axis is the cutoff degree of the polynomial $P$. The relative errors are defined by Eq.~\eqref{eq:defErrorRigorous} for the rigorous bounds and Eq.~\eqref{eq:defErrorCentral} for the central values.}
  \label{fig:error_bounds_central_feyn}
\end{figure}

As an example, we pick numerical values
\begin{equation}
  p^2 = 2, m=1 \, ,
  \label{eq:bubbleNumPoint}
\end{equation}
for bubble integrals in $d=4$, and compare our numerical results against the exact result. As $p^2$ is positive, though below the Cutkosky cut threshold $4m^2$, the Euclidean momentum space treatment in Section \ref{subsec:posMomSpace} is not applicable. SDPA-QD working at quadruple-double precision is used to compute central values and SDPA-GMP working at 8 times the double precision is used to compute rigorous bounds.
In Fig.~\ref{fig:error_bounds_central_feyn}, we plot the relative error of the central value for $\hat I^d_{2,1}$ as well as the relative error of the rigorous bounds for $\hat I^d_{2,1}$.
We can see on the log-scale plot that the numerical result again converge exponentially to the exact result as the cutoff degree is increased. In particular, with cutoff degree $N=14$, the numerical result is
\begin{equation}
  \hat I_{2,1}^{d=4} \Big|_{p^2=2, \, m=1} \approx 1.57079632679413 \, , \label{eq:bubNumericalDeg14}
\end{equation}
which is slightly smaller than the exact result, with a relative error of $4.9 \times 10^{-13}$.
\subsection{Constraints for expansions in dimensional regularization parameter}
\label{subsec:epExpansion}
The methods described in Sections \ref{subsec:posMomSpace} and \ref{subsec:posFeynSpace} are applicable to fixed spacetime dimensions, i.e.\ $d=4-2\epsilon$ with fixed values of $\epsilon$, which can be 0 if we target the 4-dimensional case, or any value larger than $(-1)$ which will preserve the ultraviolet convergence properties of the integrals involved in the text above. However, for practical applications, Feynman integrals typically need to be evaluated as a Laurent expansion in $\epsilon$. In the examples given in this paper, we can choose master integrals which are finite as $\epsilon \to 0$, so the task is to calculation their Taylor expansions in $\epsilon$. Any divergent integral can be reduced to rational-linear combinations of the master integrals, with all divergences absorbed into $\epsilon$ poles of the coefficients.\footnote{In fact, it is believed that in general, one can choose ``quasi-finite'' master integrals \cite{vonManteuffel:2014qoa} which are convergent as $\epsilon \to 0$ except for a possible $1/\epsilon$ pole that appears as an overall prefactor in the Feynman parameter representation.}

We now present two strategies for calculating the $\epsilon$ expansion, using the one-loop bubble integral example. The first strategy presented below is directly formulating positivity constraints for the $\epsilon$ expansion terms, and the second strategy presented is numerical differentiation of the results with respect to $\epsilon$ around $\epsilon = 0$.

\subsubsection{Generic constraints}
\label{subsubsec:eps_generic}
We will take a break from the bubble integrals and write down the general form of the Feynman parametrization for an $L$-loop integral with $n$ propagators,
\begin{align}
  &\quad I^d_{a_1,a_2, \dots , a_n} \equiv \left( \prod_{i=1}^L \int \frac {d^d l_i \, e^{\gamma_E \epsilon}}{i \pi^{d/2}}  \right)
  \frac{1} {\rho_1^{a_1} \rho_2^{a_2} \dots \rho_n^{a_n}} \nonumber \\
  &= \frac {\Gamma(a - L d / 2) e^{\gamma_E \epsilon}}
       {\Gamma(a_1) \Gamma(a_2) \dots \Gamma(a_n)}
  \int_{x_i \geq 0} d^n x_i \, \delta \left( 1 - \sum x_i \right) \, \left( \prod_i x_i^{a_i - 1} \right)
  \frac{\mathcal U(x_i)^{a - (L+1)d/2}} {\mathcal F(x_i)^{a - Ld/2}} \, ,
  \label{eq:generalFeynParam}
\end{align}
where $a \equiv \sum a_i$, and $\mathcal U$ and $\mathcal F$ are graph polynomials. In the Euclidean region, i.e.\ when external kinematics do not allow any Cutkosky cuts, $\mathcal U$ and $\mathcal F$ are non-negative in the range of integration. To make it easier to formulate positivity constraints, we adjust constant prefactors and slightly rewrite Eq.~\eqref{eq:generalFeynParam} as
\begin{align}
  &\quad \tilde I^d_{a_1,a_2, \dots , a_n} \equiv \frac {\Gamma(a_1) \Gamma(a_2) \dots \Gamma(a_n)} {\Gamma(a - L d / 2) e^{\gamma_E \epsilon}} I^d_{a_1,a_2, \dots , a_n} \nonumber \\
  &= \frac {\Gamma(a_1) \Gamma(a_2) \dots \Gamma(a_n)} {\Gamma(a - L d / 2) e^{\gamma_E \epsilon}} 
  \left( \prod_{i=1}^L \int \frac {d^d l_i \, e^{\gamma_E \epsilon}}{i \pi^{d/2}}  \right)
  \frac{1} {\rho_1^{a_1} \rho_2^{a_2} \dots \rho_n^{a_n}} \nonumber \\
  &= 
  \int_{x_i \geq 0} d^n x_i \, \delta \left( 1 - \sum x_i \right) \, \left( \prod_i x_i^{a_i - 1} \right)
  \frac{\mathcal U(x_i)^{a - (L+1)d/2}} {\mathcal F(x_i)^{a - Ld/2}} \, .
  \label{eq:generalFeynParam1}
\end{align}
We will restrict our attentions to values of $d$ and $a_i$ under which the RHS of Eq.~\eqref{eq:generalFeynParam1} is convergent. Since there are otherwise no restrictions on $a_i$, generally the integrals are not master integrals which are usually chosen to have small values of $a_i$.
We set
\begin{equation}
  d = d_0 - 2 \epsilon \, ,
\end{equation}
where $d_0$ is usually an integer spacetime dimension such as 4. The Taylor expansion of the LHS of Eq.~\eqref{eq:generalFeynParam1} is written as
\begin{equation}
  \tilde I_{a_1,a_2, \dots , a_n} = \tilde I_{a_1,a_2, \dots , a_n} \big|_{\epsilon^0} + \epsilon \cdot \tilde I_{a_1,a_2, \dots , a_n} \big|_{\epsilon^1} +
  \epsilon^2 \cdot \tilde I_{a_1,a_2, \dots , a_n} \big|_{\epsilon^2} \, \dots
\end{equation}
The only $\epsilon$ dependence of the RHS of Eq.~\eqref{eq:generalFeynParam1} is in the exponents on the graph polynomials, so the $\mathcal O(\epsilon^k)$ term in the Taylor expansion is
\begin{equation}
  \tilde I_{a_1,a_2, \dots , a_n} \Big|_{\epsilon^k} = 
  \int_{x_i \geq 0} d^n x_i \delta \left( 1 - \sum x_i \right) \, \left( \prod_i x_i^{a_i - 1} \right)
  \frac{\mathcal U(x_i)^{a - (L+1)d_0/2}} {\mathcal F(x_i)^{a - Ld_0/2}}
  \frac 1 {k!} \log^k \frac {\mathcal U^{L+1}} {\mathcal F^L}
  \, .
  \label{eq:epsilonExp}
\end{equation}
Note that $\mathcal U^{L+1} / \mathcal F^l$ in the equation above is a quantity that is invariant under the rescaling symmetry Eq.~\eqref{eq:projective}, since $\mathcal U$ is a homogeneous polynomial of degree $l$ and $\mathcal F$ is a homogeneous polynomial of degree $l+1$. For the Euclidean region, as $\mathcal U$ and $\mathcal F$ are positive in the range of integration, no branch-cut singularities from the logarithm are encountered. Now we write down a positivity constraint using our usual trick of constructing non-negative integrands from squares of polynomials,
\begin{equation}
  0 \leq \int_{x_i \geq 0} d^n x_i \, \delta \left( 1 - \sum x_i \right) \, \left( \prod_i x_i^{a_i - 1} \right)
  \frac{\mathcal U(x_i)^{a - (L+1)d_0/2}} {\mathcal F(x_i)^{a - Ld_0/2}}
  P^2\left( \log \frac {\mathcal U^{L+1}} {\mathcal F^L} \right)
  \, ,
  \label{eq:feynPositivity}
\end{equation}
where $P$ is an arbitrary polynomial (of the argument in the bracket) under a cutoff degree $N$, as a sum of monomials each multiplied by a free parameter. After expanding the square of the polynomial, each monomial term is identified with a term in the Taylor expansion over $\epsilon$ using Eq.~\eqref{eq:epsilonExp}. Using the same manipulations as in Sections \ref{subsec:posMomSpace} and \ref{subsec:posFeynSpace}, Eq.~\eqref{eq:feynPositivity} implies that a certain symmetric matrix is positive semidefinite. We first define an auxiliary notation
\begin{equation}
  H_k \equiv (k!) \tilde I^d_{a_1,a_2, \dots , a_n} \Big|_{\epsilon^k} \, .
\end{equation}
Then we have
\begin{align}
  \begin{pmatrix}
    H_0 & H_1 & H_2 & \dots & H_N \\
    H_1 & H_2 & H_3 & \dots & H_{N+1} \\
    H_2 & H_3 & H_4 & \dots & H_{N+2} \\
    \vdots & \vdots & \vdots & \ddots & \vdots \\
    H_N & H_{N+1} & H_{N+2} & \dots & H_{2N}
  \end{pmatrix}
  \succcurlyeq 0 \, , \label{eq:epsExpHankel}
\end{align}
where we again used the notation $\succcurlyeq 0$ to indicate that a matrix is positive semidefinite. The matrix is in a special form called a Hankel matrix, where all matrix entries are defined through a sequence $H_0$, $H_1$, \dots, $H_N$. Hankel matrices have also appeared in the context of EFT positivity bounds, e.g.\ in Ref.~\cite{Arkani-Hamed:2020blm}.

Eq.~\eqref{eq:epsExpHankel} is extremely general and applies to any convergent Feynman integral (or quasi-finite Feynman integral \cite{vonManteuffel:2014qoa} after dropping an overall divergent prefactor) in the Euclidean region with arbitrary powers of propagators and no numerators.
This tells us that the $\epsilon$ expansion of such Feynman integrals are not arbitrary but are constrained by positivity constraints which, to our best knowledge, have not been previously revealed in the literature.

\subsubsection{Taylored constraints for specific Feynman integrals}
\label{subsubsec:eps_taylored}
From Eq.~\eqref{eq:epsilonExp}, it is not hard to anticipate that more specialized positivity constraints exist if we focus on a particular family of Feynman integrals if $\log (\mathcal U^{L+1} / \mathcal F^l)$ has either an upper bound or lower bound, or both, in the range of integration. For example, for one-loop bubble integrals, the $\mathcal U$ polynomial is equal to $x_1+x_2$ and is set to 1 by the Dirac delta function in Eq.~\eqref{eq:generalFeynParam}. So we recover the Feynman parametrization for the one-loop bubble integral, Eq.~\eqref{eq:bubbleFeynParam}, with $x_1 = x, \, x_2 = 1-x, \, \mathcal U(x) = 1, \, \mathcal F(x) = m^2 - p^2x(1-x)$. We have, for $0<p^2 < 4m^2$ under consideration, in the integration range $0 \leq x \leq 1$,
\begin{equation}
  \log \frac{\mathcal U^{L+1}} {\mathcal F^L} = \log \frac 1 {\mathcal F} = \log \frac{1} {m^2 - p^2x(1-x)} \leq \log \frac{1} {m^2 - p^2 / 4} \equiv \log \max \frac{\mathcal U^{L+1}} {\mathcal F^L} \, . \label{eq:bubbleFpolyBound}
\end{equation}
Therefore
\begin{equation}
  \log \max \frac{\mathcal U^{L+1}} {\mathcal F^L} - \log \frac{\mathcal U^{L+1}} {\mathcal F^L} \label{eq:logMaxMinusLog}
\end{equation}
is a positive quantity. Similarly, if a minimum of $\log \mathcal U^{L+1} / \mathcal F^L$ exists over the range of integration, then
\begin{equation}
  \log \frac{\mathcal U^{L+1}} {\mathcal F^L} - \log \min \frac{\mathcal U^{L+1}} {\mathcal F^L} \label{eq:logMinusLogMin}
\end{equation}
is a positive quantity. In the bubble integral example, again under $0<p^2 < 4m^2$ and $0 \leq x \leq 1$,
\begin{equation}
  \log \frac{\mathcal U^{L+1}} {\mathcal F^L} = \log \frac 1 {\mathcal F} = \log \frac{1} {m^2 - p^2x(1-x)} \geq \log \frac{1} {m^2} \equiv \log \min \frac{\mathcal U^{L+1}} {\mathcal F^L} \, . \label{eq:bubbleFpolyBoundAlt}
\end{equation}

\begin{figure}
  \centering
  \includegraphics[width=0.5\textwidth]{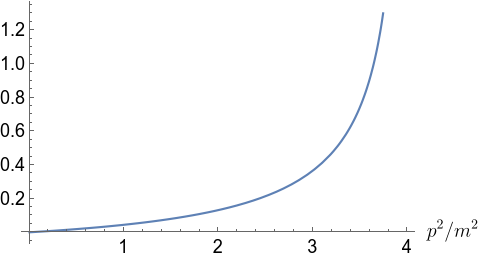}
  \caption{The RHS of Eq.~\eqref{eq:ep1_constraint_example} versus $p^2/m^2$ in the range of validity $0 < p^2 < 4$.}
  \label{fig:ep1_constraint_example}
\end{figure}

Now we show an example of using Eq.~\eqref{eq:bubbleFpolyBound} to contrain the $O(\epsilon)$ term in the expansion of the finite bubble integral $I_{2,1}$, in the style of an ad hoc constraint as was done for the $\mathcal O(\epsilon^0)$ part in Section \ref{subsubsec:crude}.\footnote{It is also possible to use Eq.~\eqref{eq:bubbleFpolyBoundAlt} instead, or in combination.} Using the definition $\hat F(x) = \mathcal F(x) / m^2$ in Eq.~\eqref{eq:bubbleHatFDef}, Eq.~\eqref{eq:bubbleFpolyBound} is rewritten as
\begin{equation}
  \log \frac 1 {\hat F(x)} \leq \log \max \frac 1 {\mathcal F} = \log \frac{1} {1 - p^2 / (4 m^2)} \, ,
\end{equation}
i.e.,
\begin{equation}
  \log \frac{1} {1 - p^2 / (4 m^2)} - \log \frac 1 {\hat F(x)} \geq 0 \, ,
  \label{eq:bubbleFpolyBoundHat}
\end{equation}
We expand both the LHS and RHS of Eq.~\eqref{eq:bubbleFeynParamInvertExample}, with $d=4 - 2\epsilon$, as a Taylor series in $\epsilon$. Equating the $\epsilon^0$ terms gives
\begin{equation}
  2 \int_0^1 dx \, x^2 \frac 1 {\hat F(x)} = \frac {p^2 - 2m^2} {2p^2} \left( \hat I_{2,1} \big|_{\epsilon^0} \right) + \frac{m^2}{p^2} \, ,
  \label{eq:bubbleFeynParamInvertExampleEp0}
\end{equation}
while equating the $\epsilon^1$ terms gives
\begin{equation}
  2 \int_0^1 dx \, x^2 \frac 1 {\hat F(x)} \cdot \log \left( \frac 1 {\hat F(x)} \right) =
  \frac {p^2 - 4m^2} {2p^2} \left( \hat I_{2,1} \big|_{\epsilon^0} \right)
  + \frac {p^2 - 2m^2} {2p^2} \left( \hat I_{2,1} \big|_{\epsilon^1} \right)
  + \frac{2m^2}{p^2} \, .
  \label{eq:bubbleFeynParamInvertExampleEp1}
\end{equation}
Now we use Eq.~\eqref{eq:bubbleFpolyBoundHat} to write down the positivity constraint
\begin{equation}
  2 \int_0^1 dx \, x^2 \frac 1 {\hat F(x)} \left( \log \frac{1} {1 - p^2 / (4 m^2)} - \log \frac 1 {\hat F(x)} \right) \geq 0 \, .
\end{equation}
Then applying Eqs.~\eqref{eq:bubbleFeynParamInvertExampleEp0} and \eqref{eq:bubbleFeynParamInvertExampleEp1} leads to a constraint on $\hat I_{2,1} \big|_{\epsilon^1}$,
\begin{align}
  0 & \leq \left( \frac {p^2-2} {2p^2} \log \frac{1} {1 - p^2 / (4 m^2)} + \frac {4m^2-p^2}{2p^2} \right) \left( \hat I_{2,1} \big|_{\epsilon^0} \right) + \frac{2m^2 - p^2}{2p^2} \left( \hat I_{2,1} \big|_{\epsilon^1} \right) \nonumber \\
  & + \left( \log \frac{1} {1 - p^2 / (4 m^2)} - 2 \right) \frac {m^2} {p^2} \, .
  \label{eq:ep1_constraint_example}
\end{align}
We plot the RHS of Eq.~\eqref{eq:ep1_constraint_example} versus $p^2/m^2$ in Fig.~\ref{fig:ep1_constraint_example}. We can see that it is indeed non-negative in the range $0 < p^2/m^2 < 4$.
\subsubsection{Bubble integral up to second order in $\epsilon$}
\label{sec:bubbleResultsEps}
We proceed to combine positivity constraints for $\epsilon$ expansion coefficients covered in Sections \ref{subsubsec:eps_generic} and \ref{subsubsec:eps_taylored} with the semidefinite programming technique already used in Sections \ref{subsec:posMomSpace} and \ref{subsec:posFeynSpace}, in order to obtain high-precision results for the $\epsilon$ expansion of the bubble integral.

Recall that we evaluated the $\mathcal O(\epsilon^0)$ part, i.e.\ the $d=4$ result, for the bubble integral starting from Eq.~\eqref{eq:posAnsatzFeynParam} and the refined version Eq.~\eqref{eq:bubbleFeynRefinedPositivity}. For simplicity, we will build upon the first version, Eq.~\eqref{eq:posAnsatzFeynParam}, and use the additional positive building block Eq.~\eqref{eq:bubbleFpolyBoundHat} to write down the following positivity constraint at $d=d_0-2 \epsilon = 4-2 \epsilon$,
\begin{align}
  0 & \leq \int_0^1 dx \, x P(x)^2 \frac 1 {\hat F(x)^{3-d_0/2}}
  \left( \log \max \frac{1} {\hat F} - \log \frac{1} {\hat F(x)} \right) \\
  &=
  \int_0^1 dx \, x P(x)^2 \frac 1 {[1 - x(1-x) p^2/m^2]}
  \left( \log \frac{1} {1 - p^2/(4m^2)} - \log \frac{1} {1 - x(1-x) p^2/m^2} \right)
  \, ,
  \label{eq:posAnsatzFeynParamEp1}
\end{align}
where $P(x)$ is again an arbitrary polynomial in $x$, and we are free to choose the maximum degree of monomials that are included, depending on the accuracy we would like to attain. After expanding $P(x)^2$ into a sum of monomials, the contribution from each monomial can be evaluated following the same procedure as used in the example in Section \ref{subsubsec:eps_taylored}.
In particular, the result will be a sum of terms proportional to $\hat I_{2,1} \big|_{\epsilon^0}$, terms proportional to $\hat I_{2,1} \big|_{\epsilon^1}$, and constant terms. The remaining calculation steps are very similar to those of Section \ref{subsec:posFeynSpace}. Re-using the parametrization Eq.~\eqref{eq:PxParametrization} for the polynomial $P$ with a cutoff degree $N=14$, we again arrive at
\begin{equation}
  {\vec \alpha}^T \, \mathbb M \, \vec \alpha \geq 0 \, ,
\end{equation}
stating that an appropriate matrix $\mathbb M$ is positive semidefinite, i.e.
\begin{equation}
  \mathbb M \succcurlyeq 0 \, .
\end{equation}
In this case, $\mathbb M$ is a sum of three terms,
\begin{equation}
  \mathbb M = \mathbb M_1 + I_{2,1}\big|_{\epsilon^0} \cdot \mathbb M_2 + I_{2,1}\big|_{\epsilon^1} \cdot \mathbb M_3 \, . \label{eq:matM_sum_three_parts}
\end{equation}
where the three matrices $\mathbb M_{1,2,3}$, with rational dependence on $p^2$ and $m$, are obtained from dimension shifting, IBP, and finally $\epsilon$-expansion as in the example of Section \ref{subsubsec:eps_taylored}. We approximate $I_{2,1}\big|_{\epsilon^0}$ to be the central value Eq.~\eqref{eq:bubNumericalDeg14} obtained in the previous $d=4$ calculation with the same cutoff degree $N=14$. At this point, $I_{2,1}\big|_{\epsilon^1}$ remains the only unknown parameter on the RHS of Eq.~\eqref{eq:matM_sum_three_parts}, and its allowed range as well as the central value can be determined by semidefinite programming solvers as covered in previous sections. We again use SDPA-QD to produce the ``central value'' for the numerical result, defined by the same prescription as before, obtaining
\begin{equation}
  I_{2,1}\big|_{\epsilon^1} \approx 0.74313814320586 \, , \ \text{at } {p^2=2, m=1} \, ,
\end{equation}
which is slightly larger than the exact result with a relative error of $4.3 \times 10^{-12}$.

We continue to present how the $\mathcal O(\epsilon^2)$ term of the bubble integral is calculated. We use the positivity constraint,
\begin{equation}
  0 \leq \int_0^1 dx \, x P \left( x, \log \frac{1} {\hat F} \right)^2 \left[ m^2 - p^2 x (1-x) \right]^{d_0/2 - 3}  \, ,
  \label{eq:posAnsatzFeynParamEp2}
\end{equation}
where $P$ is a polynomial in $x$ and $\log 1 / \hat F(x)$ with maximum degrees $N_1$ and $N_2$ in the two variables, parametrized as
\begin{equation}
  P\left( x, \log \frac{1} {\hat F(x)} \right) = \sum\limits_{0\leq i_1 \leq N_1} \sum\limits_{0 \leq i_2 \leq N_2} \alpha_{i_1, i_2} \, x^{i_1} \left( \log \frac{1} {\hat F(x)} \right)^{i_2} \, .
\end{equation}
We use $N_1=14$ and $N_2 = 1$, so that there are at most 14 power of $x$ in $P(x)$ and at most one power of $\log [1 / \hat F(x)]$. After expanding the $P^2$, there are at most two powers of the aforementioned logarithm, therefore there are $\epsilon$ expansions coefficients at orders $\epsilon^0$, $\epsilon^1$, and $\epsilon^2$. We obtain an expression of the form
\begin{equation}
  {\vec \alpha}^T \, \mathbb M \, \vec \alpha \geq 0 \, ,
\end{equation}
where the column vector $\vec \alpha$ groups together all the $\alpha_{i_1, i_2}$ parameters. The matrix $\mathbb M$ is the sum of a constant term, a term proportional to $I_{2,1}\big|_{\epsilon^0}$, a term proportional to $I_{2,1}\big|_{\epsilon^1}$, and finally a term proportional to $I_{2,1}\big|_{\epsilon^2}$. We use previous numerical results for $I_{2,1}\big|_{\epsilon^0}$ and $I_{2,1}\big|_{\epsilon^1}$, and solve a semidefinite programming problem to find the central value of $I_{2,1}\big|_{\epsilon^2}$ to be
\begin{equation}
  I_{2,1}\big|_{\epsilon^2} \approx 0.208108744452 \, , \ \text{at } {p^2=2, m=1} \, ,
\end{equation}
which is slightly larger than the exact result with a relative error of $1.9 \times 10^{-11}$.

There is no obstruction to obtaining results at even higher orders in the $\epsilon$ expansion for the bubble integral example. Generally, we calculate iteratively to higher and higher orders in $\epsilon$, at each step taking previous numerical results as known input. The positivity constraint is Eq.~\eqref{eq:posAnsatzFeynParamEp2} with an appropriate cutoff degree for $\log [1 / \hat F(x)]$ depending on the desired order in the $\epsilon$ expansion. The RHS of Eq.~\eqref{eq:posAnsatzFeynParamEp2} can be optionally multiplied by either Eq.~\eqref{eq:logMaxMinusLog} or Eq.~\eqref{eq:logMinusLogMin}, with $\mathcal U=1$ and $L=1$ in the case of bubble integrals, to give more constraints. This will produce constraints for bubble integrals to any desired order in the $\epsilon$ expansion.
\subsection{$\epsilon$ expansion from numerical differentiation w.r.t.\ spacetime dimension}
\label{subsec:numericalDiff}
Here we present an alternative method for numerically evaluating the $\epsilon$-expansion of the normalized bubble integral $\hat I^d_{2,1}$ defined in Eq.~\eqref{eq:bubbleMasterParam}. Instead of formulating constraints for the $\epsilon$ expansion coefficients, we calculate the $\hat I^d_{2,1}$ at numerical values of the spacetime dimension near 4, and use finite-difference approximations to obtain derivatives w.r.t.\ $\epsilon$. The derivatives are related to the terms in the $\epsilon$ expansion via
\begin{equation}
  \hat I_{2,1} \big|_{\epsilon^k} = \frac{1} {k!} \frac{d^k} {d \epsilon^k} \hat I^{d=4-2\epsilon}_{2,1} \Big|_{\epsilon=0} \, .
\end{equation}
For an arbitrary function $f(\epsilon)$, we use 4th-order finite-difference approximations,
\begin{align}
  \frac{d} {d \epsilon} f(\epsilon) \bigg|_{\epsilon = \epsilon^0} & \approx
  \frac 1 {\Delta \epsilon} \bigg[ \frac 1 {12} f(\epsilon_0 - 2 \Delta \epsilon) - \frac 2 {3} f(\epsilon_0 - \Delta \epsilon) \nonumber \\
    & \qquad + \frac 2 {3} f(\epsilon_0 + \Delta \epsilon) - \frac 1 {12} f(\epsilon_0 + 2 \Delta \epsilon) \bigg] \, , \label{eq:finiteDiff1} \\
  \frac {d^2} {d \epsilon^2} f(\epsilon) \bigg|_{\epsilon = \epsilon^0} & \approx
  \frac 1 {\Delta \epsilon^2} \bigg[ -\frac 1 {12} f(\epsilon_0 - 2 \Delta \epsilon) + \frac 4 {3} f(\epsilon_0 - \Delta \epsilon) - \frac 5 2 f(\epsilon_0) \nonumber \\
    &\qquad + \frac 4 {3} f(\epsilon_0 + \Delta \epsilon) - \frac 1 {12} f(\epsilon_0 + 2 \Delta \epsilon) \bigg] \, , \label{eq:finiteDiff2}
\end{align}
where $\Delta \epsilon$ is the step size. As the name suggests, these formulas are exact when $f$ is a polynomial with a degree up to 4.
Note that the method of Section \ref{subsec:posFeynSpace} can be used to evaluate the normalized bubble integral $\hat I^d_{2,1}$ in any spacetime dimension $d<6$, i.e. $\epsilon > -1$, so Eqs.~\eqref{eq:finiteDiff1} and \eqref{eq:finiteDiff2} can be readily used with $\epsilon_0 = 0$ and a small $\Delta \epsilon$, chosen to be
\begin{equation}
  \Delta \epsilon = 10^{-3} \, .
\end{equation}
We again choose kinematic parameter values $p^2=2$ and $m=1$. The final results from numerical differentiation, up to the second order in $\epsilon$, are
\begin{equation}
  \hat I_{2,1} \big|_{\epsilon^1} \approx 0.7431381432049 \, ,
\end{equation}
which is slightly larger than the exact result with a relative error of $3.2 \times 10^{-12}$, and
\begin{equation}
  \hat I_{2,1} \big|_{\epsilon^2} \approx 0.20810874450 \, ,
\end{equation}
which is slightly larger than the exact result with a relative error of $2.8 \times 10^{-10}$.

\section{Three-loop banana integrals with unequal masses}
\label{sec:banana}
\subsection{Definitions and conventions}
\label{subsec:bananaDefinitions}
\begin{figure}
  \centering
  \includegraphics[width=0.3\textwidth]{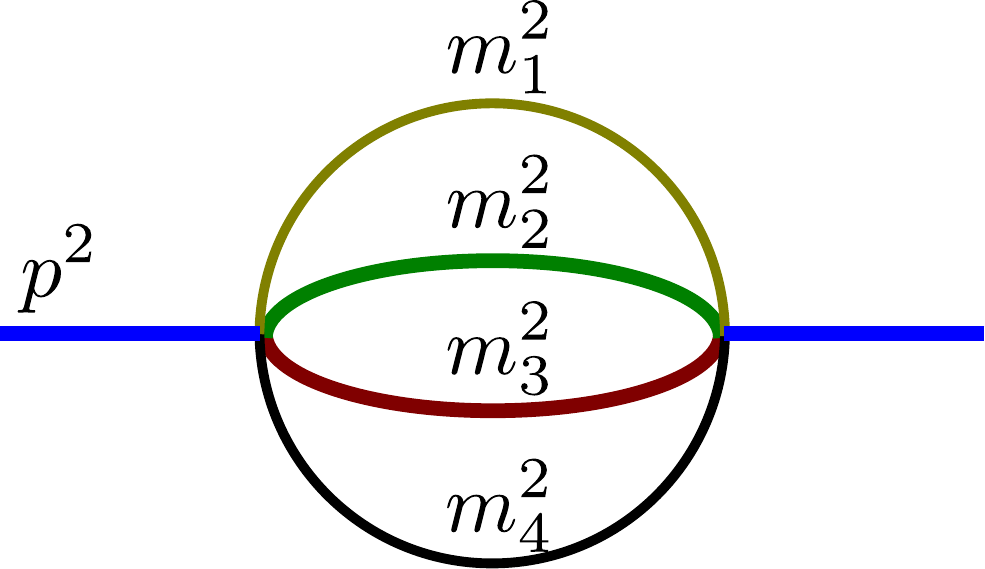}
  \caption{Three-loop banana family of integrals, with internal and external squared masses labeled.}
  \label{fig:banana}
\end{figure}

Here we present a three-loop example, the so called banana integrals. Banana integrals are various loop orders have received intense interest from an analytic perspective due to connections with Calabi-Yau manifold \cite{Groote:2005ay, Klemm:2019dbm, Bonisch:2020qmm, Bonisch:2021yfw, Kreimer:2022fxm, Forum:2022lpz, Pogel:2022ken}. We apply our numerical method developed in the previous section \ref{sec:bubble} on one-loop bubble integrals, with some minor adaptations, to evaluate $11$ nontrivial master integrals for the banana diagram. We assume the readers to be familiar with the previous section as many shared techniques will not be introduced again.
The diagram for the integrals is shown in Fig.~\ref{fig:banana}.
Due to dimension-shifting identities reviewed in Section \ref{subsec:bubbleIBP}, the $\epsilon$ expansions of integrals in $d=4-2\epsilon$ and $d=2-2\epsilon$ can be related to each other. We will always use
\begin{equation}
  d=2-2\epsilon \, ,
\end{equation}
for three-loop banana integrals, which is convenient as the scalar integral has no ultraviolet divergence in this spacetime dimension.
The banana family of integrals is defined as
\begin{align}
  I_{a_1, a_2, a_3, a_4} &\equiv \left( \prod_{i=1}^3 \int \frac {d^d l_i \, e^{\gamma_E \epsilon}} {i \pi^{d/2}} \right)
  \frac 1 {(-l_1^2 + m_1^2)^{a_1}}
  \frac 1 {(-l_2^2 + m_2^2)^{a_2}}
  \frac 1 {(-l_3^2 + m_3^2)^{a_3}} \nonumber \\
  &\qquad \times 
  \frac 1 {[-(p+l_1+l_2+l_3)^2 + m_4^2]^{a_4}}\, , \qquad \text{with } d = 2 - 2\epsilon \, . \label{eq:banana}
\end{align}
We have suppressed the $d$ dependence on the LHS of the above equation, unlike the one-loop case Eq.~\eqref{eq:selfEnergyInt}, since we will not make use of dimension-shifting in the treatment of three-loop banana integrals and will exclusively work with $d= 2 - 2\epsilon$.
If any one of the four indices $a_i$ is non-positive in Eq.~\eqref{eq:banana}, the remaining propagators have the structure of three one-loop massive tadpole integrals. For example, if $a_4=0$, Eq.~\eqref{eq:banana} clearly factorizes into the product of three scalar tadpole integrals. For facilitating the discussion of positivity constraints, it will be convenient to define a variant of Eq.~\eqref{eq:banana} with slightly adjusted constant factors,
\begin{align}
  \hat I_{a_1, a_2, a_3, a_4} &\equiv \left( \prod_{i=1}^3 \int \frac {d^d l_i} {i \pi^{d/2}} \right) \frac {1} {\Gamma(4 - 3d/2)}
  \frac 1 {(-l_1^2 + m_1^2)^{a_1}}
  \frac 1 {(-l_2^2 + m_2^2)^{a_2}}
  \frac 1 {(-l_3^2 + m_3^2)^{a_3}} \nonumber \\
  &\qquad \times 
  \frac 1 {[-(p+l_1+l_2+l_3)^2 + m_4^2]^{a_4}} \, . \label{eq:bananaHat}
\end{align}

By IBP reduction, all integrals of the banana family, with integer values of $a_i$ in Eq.~\eqref{eq:bananaHat}, can be expressed as linear sums of 15 master integrals. Publicly available software, such as those presented Refs.~\cite{Anastasiou:2004vj, vonManteuffel:2012np, Smirnov:2008iw, Smirnov:2019qkx, Lee:2012cn, Lee:2013mka, Maierhofer:2017gsa}, can be used to give a list of master integrals as well as performing the actual IBP reduction of integrals. There are 11 nontrivial ``top-level'' master integrals that are not products of tadpole integrals, shown in three groups below according to the total number of the four indices,
\begin{equation}
  \begin{aligned}
    &\hat I_{1,1,2,2}, \, \hat I_{1,2,1,2}, \, \hat I_{1,2,2,1}, \, \hat I_{2,1,1,2}, \, \hat I_{2,1,2,1}, \, \hat I_{2,2,1,1}, \\
    &\hat I_{1,1,1,2}, \, \hat I_{1,1,2,1}, \, \hat I_{1,2,1,1}, \, \hat I_{2,1,1,1}, \\
    &\hat I_{1,1,1,1} \, .
  \end{aligned}
  \label{eq:mastersBananaTop}
\end{equation}
In addition, there are 4 master integrals that are trivial products of tadpole integrals. These master integrals are chosen as
\begin{equation}
  \hat I_{0,2,2,2}, \, \hat I_{2,0,2,2}, \, \hat I_{2,2,0,2}, \, \hat I_{2,2,2,0} \, , \label{eq:mastersTadCubed}
\end{equation}
where we raised every propagator to a 2nd power to make the integral UV finite in $d=2-2\epsilon$. Each of these four master integrals is a product of three one-loop tadpole integrals given in Eq.~\eqref{eq:tadpoleGeneral} with $n=2$, with some adjustment of the overall factor according to Eq.~\eqref{eq:bananaHat},
\begin{align}
  & \hat I_{0,2,2,2} = \frac {\Gamma^3(1+\epsilon)}{\Gamma(1+3\epsilon)} \left( \frac 1 {m_2^2 m_3^2 m_4^2} \right)^{1+\epsilon} , \quad
  \hat I_{2,0,2,2} = \frac {\Gamma^3(1+\epsilon)}{\Gamma(1+3\epsilon)} \left( \frac 1 {m_1^2 m_3^2 m_4^2} \right)^{1+\epsilon} , \nonumber \\
  & \hat I_{2,2,0,2} = \frac {\Gamma^3(1+\epsilon)}{\Gamma(1+3\epsilon)} \left( \frac 1 {m_1^2 m_2^2 m_4^2} \right)^{1+\epsilon} , \quad
  \hat I_{2,2,2,0} = \frac {\Gamma^3(1+\epsilon)}{\Gamma(1+3\epsilon)} \left( \frac 1 {m_1^2 m_2^2 m_3^2} \right)^{1+\epsilon} \, .
  \label{eq:tadpoleCubedValues}
\end{align}
The values of the 11 remaining master integrals in Eq.~\eqref{eq:mastersBananaTop} will be calculated numerically from positivity constraints. We will work with kinematic variables in the range
\begin{equation}
  p^2 < (m_1+m_2+m_3+m_4)^2 \, ,
  \label{eq:bananaEucRegion}
\end{equation}
i.e.\ below the particle production threshold, which ensures that the integrals are real-valued.
Similar to the case of one-loop bubble integrals, when the chosen value of $p^2$ is non-negative, we cannot embed the integrals into Euclidean momentum space and need to use Feynman-parameter space to formulate positivity constraints. The Feynman parametrization follows from the general formula Eq.~\eqref{eq:generalFeynParam} with adjustment of constant factors according to Eq.~\eqref{eq:bananaHat},
\begin{align}
  \hat I_{a_1, a_2, a_3, a_4} &= \frac {\Gamma(a-3d/2) / \Gamma(4-3d/2)} {\Gamma(a_1) \Gamma(a_2) \Gamma(a_3) \Gamma(a_4)} \int_{x_i \geq 0} dx_1 dx_2 dx_3 dx_4 \, \delta(1-x_1-x_2-x_3-x_4) \nonumber \\
  &\quad \times \left( \prod_{i=1}^4 x_i^{a_i-1} \right) \frac {\mathcal U(x_i)^{a-2d}} {\mathcal F(x_i)^{a-3d/2}} \, ,
  \label{eq:feynParamBanana}
\end{align}
where we used the definition $a \equiv a_1+a_2+a_3+a_4$. The two graph polynomials $\mathcal U$ and $\mathcal F$ are, for the banana family of integrals,
\begin{align}
  \mathcal U(x_1, x_2, x_3, x_4) &= x_2 x_3 x_4 + x_1 x_3 x_4 + x_1 x_2 x_4 + x_1 x_2 x_3 , \nonumber \\
  \mathcal F(x_1, x_2, x_3, x_4) &= p^2 x_1 x_2 x_3 x_4 + (m_1^2 x_1 + m_2^2 x_2 + m_3^2 x_3 + m_4^2 x_4) \mathcal U(x_1, x_2, x_3, x_4) \, .
\end{align}
Note that we have
\begin{equation}
  \mathcal U(x_1, x_2, x_3, x_4) \geq 0, \quad \mathcal F(x_1, x_2, x_3, x_4) \geq 0 \, ,
\end{equation}
in the integration region of Eq.~\eqref{eq:feynParamBanana}, i.e.\ $x_i \geq 0, \, \sum_i x_i=1$. This will help us formulate positivity constraints. As in the one-loop bubble case, except for the ``gauge fixing'' Dirac delta function, the rest of Eq.~\eqref{eq:feynParamBanana} has the projective invariance Eq.~\eqref{eq:projective}, as the $\mathcal U$ and $\mathcal F$ polynomials are homogeneously of degree 3 and 4, respectively.

\subsection{Positivity constraints}
We define $\tilde x_i$ variables 
\begin{equation}
  \tilde x_i = \frac{\mathcal U(x_1, x_2, x_3, x_4)} {\mathcal F(x_1, x_2, x_3, x_4)} x_i, \quad i=1,2,3,4 \, ,
  \label{eq:rescaleXi}
\end{equation}
which are invariant under the scaling Eq.~\eqref{eq:projective}. We rewrite Eq.~\eqref{eq:feynParamBanana} as
\begin{align}
  &\quad \frac {\Gamma(4-3d/2)} {\Gamma(a-3d/2)} \Gamma(a_1) \Gamma(a_2) \Gamma(a_3) \Gamma(a_4) \, \hat I_{a_1, a_2, a_3, a_4} \nonumber \\
  &= \int_{x_i \geq 0} dx_1 dx_2 dx_3 dx_4 \, \delta(1-x_1-x_2-x_3-x_4) \left( \prod_{i=1}^4 \tilde x_i^{a_i-1} \right) \frac {\mathcal U(x_i)^{4-2d}} {\mathcal F(x_i)^{4-3d/2}} \, ,
  \label{eq:feynParamRescaledXi}
\end{align}
again with $a \equiv a_1+a_2+a_3+a_4$. Note that if $a_i \geq 1$, $a \geq 4$,
\begin{equation}
  \frac {\Gamma(4-3d/2)} {\Gamma(a-3d/2)} = \frac 1 {(4-3d/2) (5-3d/2) \dots (a-1-3d/2)}
\end{equation}
is a rational function in $d$.
With any non-negative polynomial $Q(\tilde x_i)$, we formulate a positivity constraint,
\begin{align}
  0 &\leq \int_{x_i \geq 0} dx_1 dx_2 dx_3 dx_4 \, \delta(1-x_1-x_2-x_3-x_4) Q(\tilde x_i) \frac {\mathcal U(x_i)^{4-2d}} {\mathcal F(x_i)^{4-3d/2}} \, ,
  \label{eq:bananaPos}
\end{align}
which is compatible with the projective invariance Eq.~\eqref{eq:projective}.
After expanding the polynomial $Q(\tilde x_i)$ into a sum of monomials, the contribution of each monomial $\prod_i \tilde x_i^{a_i-1}$ can be written as some $\hat I_{a_1, a_2, a_3, a_4}$ multiplied by a prefactor that is rational in $d$, according to Eq.~\eqref{eq:feynParamRescaledXi}. All such integrals are UV convergent by power counting and also IR convergent due to internal masses. No change of spacetime dimensions is involved, unlike the treatment of one-loop bubble integrals in Section \ref{subsec:posFeynSpace}.

We consider the following choices of $Q(\tilde x_i)$, with the help of an arbitrary polynomial $P(\tilde x_i)$ under a chosen maximum degree,
\begin{align}
\text{choice 1:} \quad Q(\tilde x_i) &= P(\tilde x_i)^2 , \label{eq:Qchoice1} \\
\text{choice 2:} \quad Q(\tilde x_i) &= \tilde x_1 P(\tilde x_i)^2 , \label{eq:Qchoice2} \\
\text{choice 3:} \quad Q(\tilde x_i) &= \tilde x_2 P(\tilde x_i)^2 , \\
\text{choice 4:} \quad Q(\tilde x_i) &= \tilde x_3 P(\tilde x_i)^2 , \\
\text{choice 5:} \quad Q(\tilde x_i) &= \tilde x_4 P(\tilde x_i)^2 \, . \label{eq:Qchoice5}
\end{align}
With any of the above five choices for $Q(\tilde x_i)$ and with any choice of $P(\tilde x_i)$, the inequality Eq.~\eqref{eq:bananaPos} must hold. The general form of $P(\tilde x_i)$ is a sum of all monomials under a chosen cutoff degree, each multiplied by an arbitrary coefficient. For example, if the cutoff degree is 1, then $P(\tilde x_i)$ is parametrized as
\begin{equation}
  \text{cutoff degree 1:} \quad P(\tilde x_i) = \alpha_{0,0,0,0} + \alpha_{1,0,0,0} \tilde x_1 + \alpha_{0,1,0,0} \tilde x_2 + \alpha_{0,0,1,0} \tilde x_3 + \alpha_{0,0,0,1} \tilde x_4 \, .
\end{equation}
With cutoff degree $N$, the parametrization is
\begin{equation}
  \text{cutoff degree $N$:} \quad P(\tilde x_i) = \sum_{i_1, i_2, i_3, i_4 \geq 0}^{i_1+i_2+i_3+i_4 \leq N} \alpha_{i_1, i_2, i_3, i_4} \tilde x_1^{i_1} \tilde x_2^{i_2} \tilde x_3^{i_3} \tilde x_4^{i_4} \, ,
  \label{eq:bananaGeneralP}
\end{equation}
where the number of free parameters $\alpha_{i_1, i_2, i_3, i_4}$ is equal to ${N+4 \choose 4}$ by combinatorics.

Since we have already discussed how to set up semidefinite optimization programs in the context of one-loop bubble integrals, we will be brief in covering the analogous steps here. Grouping the $\alpha_{i_1, i_2, i_3, i_4}$ parameters into a column vector $\vec \alpha$ of length ${N+4 \choose 4}$, the five choices of $Q$, Eqs.~\eqref{eq:Qchoice1} to \eqref{eq:Qchoice5}, lead to
\begin{align}
  & (\vec \alpha)^T \mathbb M^{(1)} \vec \alpha \geq 0 , \quad
  (\vec \alpha)^T \mathbb M^{(2)} \vec \alpha \geq 0 , \quad 
  (\vec \alpha)^T \mathbb M^{(3)} \vec \alpha \geq 0 , \nonumber \\
  & (\vec \alpha)^T \mathbb M^{(4)} \vec \alpha \geq 0 , \quad
  (\vec \alpha)^T \mathbb M^{(5)} \vec \alpha \geq 0 ,
\end{align}
respectively, for any values of the vector $\alpha$. For reasons we do not fully understand, the first constraint $(\vec \alpha)^T \mathbb M^{(1)} \vec \alpha \geq 0$ leads to poor numerical convergence and is discarded. The remaining four constraints are rewritten as requiring the matrices to be positive semidefinite using the notation Eq.\eqref{eq:posGeq},
\begin{equation}
  \mathbb M^{(2)} \succcurlyeq 0, \quad \mathbb M^{(3)} \succcurlyeq 0, \quad \mathbb M^{(4)} \succcurlyeq 0, \quad \mathbb M^{(5)} \succcurlyeq 0 \, .
\end{equation}
For convenience, this can be rephrased as the positive semidefiniteness of a single matrix which contains the above four matrices as diagonal blocks,
\begin{equation}
  \mathbb M = 
  \begin{pmatrix}
    \mathbb M^{(2)} & 0 & 0 & 0 \\
    0 & \mathbb M^{(3)} & 0 & 0 \\
    0 & 0 & \mathbb M^{(4)} & 0 \\
    0 & 0 & 0 & \mathbb M^{(5)} \\
  \end{pmatrix}
  \succcurlyeq 0 \, . \label{eq:bananaBlockMatPos}
\end{equation}

Analogous to the case of one-loop bubble integrals, IBP reduction expresses $\mathbb M$ as a linear combination of the 15 master integrals in Eq.~\eqref{eq:mastersBananaTop} and \eqref{eq:mastersTadCubed}, each multiplied by a matrix of rational functions in $p^2, m_1^2, m_2^2, m_3^2, m_4^2$. It is necessary to perform IBP reduction for banana integrals with up to 13 additional powers of propagators (beyond the standard 1st power), since the positive polynomial $Q$ in Eqs.~\eqref{eq:Qchoice2}-\eqref{eq:Qchoice5} bring 13 powers of $x_i$ when $P$ has degree 6. The four master integrals in Eq.~\eqref{eq:mastersTadCubed} are known analytically in Eq.~\eqref{eq:tadpoleCubedValues}, and the values of the remaining 11 master integrals are unknown parameters to be constrained by Eq.~\eqref{eq:bananaBlockMatPos}.

Before presenting numerical results, we also formulate positivity constraints for the $\epsilon$ expansion of banana integrals. Recall that banana integrals in $d=2-2\epsilon$, normalized according to Eq.~\eqref{eq:bananaHat}, has a Feynman-parameter representation Eq.~\eqref{eq:feynParamRescaledXi} using redefined Feynman parameters in Eq.~\eqref{eq:rescaleXi}. For any integers $a_i \geq 1$, Taylor-expanding both sides of Eq.~\eqref{eq:feynParamRescaledXi} and equating the coefficients of the $\epsilon^k$ term for any integer $k$, we have
\begin{equation}
  \begin{aligned}
    &\quad \left[ \frac {\Gamma(4-3d/2)} {\Gamma(a-3d/2)} \Gamma(a_1) \Gamma(a_2) \Gamma(a_3) \Gamma(a_4) \, \hat I_{a_1, a_2, a_3, a_4} \right] \Bigg |_{\epsilon^k} \\
    &= \int_{x_i \geq 0} dx_1 dx_2 dx_3 dx_4 \, \delta(1-x_1-x_2-x_3-x_4) \left( \prod_{i=1}^4 \tilde x_i^{a_i-1} \right) \frac {1} {\mathcal F(x_i)} \\
    & \quad \times \frac 1 {k!}  \log^k \frac {\mathcal U^4(x_i)} {\mathcal F^3(x_i)} \, .
  \end{aligned}
  \label{eq:feynParamRescaledXiEps}
\end{equation}
By integration-by-parts reduction, the LHS of the above equation can be written as linear combinations of the $\epsilon$ expansions of the 15 master integrals up to the $\epsilon^k$ order, assuming that the coefficients of the master integrals (from IBP reduction) are finite as $\epsilon \to 0$, which is the case here. Now we are ready to write down positivity constraints for the $\epsilon$ expansion by extending Eq.~\eqref{eq:bananaPos},
\begin{align}
  0 &\leq \int_{x_i \geq 0} dx_1 dx_2 dx_3 dx_4 \, \delta(1-x_1-x_2-x_3-x_4) Q\left( \tilde x_i, \log \frac {\mathcal U^4(x_i)} {\mathcal F^3(x_i)} \right) \frac {\mathcal U(x_i)^{4-2d}} {\mathcal F(x_i)^{4-3d/2}} \, ,
  \label{eq:bananaPosEps}
\end{align}
where $Q$ is now a positive polynomial in its two arguments above.
To build the most general form of $Q$, we follow Section \ref{subsubsec:eps_taylored} and use the building block
\begin{equation}
  \log \max \frac {\mathcal U^4} {\mathcal F^3} - \log \frac {\mathcal U^4(x_i)} {\mathcal F^3(x_i)} \geq 0 \, .
  \label{eq:bananaMaxUFBlock}
\end{equation}
The value of $\max (\mathcal U^4 / \mathcal F^3)$ will be found numerically once $p^2$ and $m_i^2$ parameters are specified in Eq.~\eqref{eq:bananaNumericPoint}, in the next subsection on numerical results. However, we find no minimum of $\log (\mathcal U^4/\mathcal F^3)$ at the same parameter values, as $\mathcal U^4 / \mathcal F^3$ can become arbitrarily close to zero (from above) in the range of integration.
We will use the following choices of $Q$,
\begin{align}
  & Q \left( \tilde x_i, \log \frac {\mathcal U^4(x_i)} {\mathcal F^3(x_i)} \right) = \tilde x_k P^2 \left( \tilde x_i, \log \frac {\mathcal U^4(x_i)} {\mathcal F^3(x_i)} \right) , \label{eq:QepsChoice1} \\
  \text{or } \ & Q \left( \tilde x_i, \log \frac {\mathcal U^4(x_i)} {\mathcal F^3(x_i)} \right) = \tilde x_k \left( \log \max \frac {\mathcal U^4} {\mathcal F^3} - \log \frac {\mathcal U^4(x_i)} {\mathcal F^3(x_i)} \right) P^2 \left( \tilde x_i, \log \frac {\mathcal U^4(x_i)} {\mathcal F^3(x_i)} \right) \, , \label{eq:QepsChoice2}
\end{align}
where $k$ can be 1, 2, 3, or 4, and $P$ is an arbitrary polynomial with a maximum total degree $N_1$ for the four $\tilde x_i$ variables and maximum degree $N_2$ for $\log (\mathcal U^4/\mathcal F^3)$.
Similar to the bubble integral case in Section \ref{subsec:epExpansion}, to constrain the $\mathcal O(\epsilon)$ part of the master integrals, we will use Eq.~\eqref{eq:QepsChoice2} with $N_2=0$, and to constrain the $\mathcal O(\epsilon^2)$ part, we will use Eq.~\eqref{eq:QepsChoice1} $N_2=1$. For both $\mathcal O(\epsilon^1)$ and $\mathcal O(\epsilon^2)$ parts, $N_1$ will be chosen to be the same as the cutoff degree used for the $\mathcal O(\epsilon^0)$ calculation.

\subsection{Numerical results}
We present numerical results for the 11 nontrivial master integrals of the banana family in Eq.~\eqref{eq:mastersBananaTop} at the following numerical values for kinematic variables,
\begin{equation}
  p^2 = 2, \quad m_1^2 = 2, \quad m_2^2 = 3/2, \quad m_3^2 = 4/3, \quad m_4^2 = 1 \, .
  \label{eq:bananaNumericPoint}
\end{equation}
We remind readers that the spacetime dimension is set to
\begin{equation}
  d = 2 - 2\epsilon \, .
\end{equation}
As this paper is aimed at illustrating a new method, we have chosen example integrals that are known to high precision in the existing literature. For three-loop banana integrals, high precision results from series solutions to differential equations are available from the DiffExp package \cite{Hidding:2020ytt}. In fact, we have chosen the same values for the masses in Eq.~\eqref{eq:bananaNumericPoint} as the example in the aforementioned paper, though we chose a different value of $p^2$ as we restrict to the Euclidean region Eq.~\eqref{eq:bananaEucRegion}. DiffExp is used to compute the master integrals at the kinematic point Eq.~\eqref{eq:bananaNumericPoint} by solving ordinary differential equations along a line in kinematic space parametrized by a variable $x$,
\begin{equation}
  p^2 = 2, \quad m_1^2 = 1+x, \quad m_2^2 = 1+x/2, \quad m_3^2 = 1+x/3, \quad m_4^2 = 1 \, , \label{eq:bananaDELine}
\end{equation}
subject to appropriate boundary conditions. The values of the master integrals at $x=1$ are then the ones needed at the point Eq.~\eqref{eq:bananaNumericPoint}.
A precision of about $10^{-54}$ is reached, and the results are effectively taken as exact values for the purpose of validating our numerical results.

For the values of the master integrals at $\mathcal O(\epsilon^0)$, i.e.\ in exactly $d=2$, we will use cutoff degrees of up to 6 in Eq.~\eqref{eq:bananaGeneralP}. Unlike the one-loop case in Section \ref{sec:bubble}, we will not fully characterize the allowed parameter region which is a sub-region of an 11-dimensional parameter space and cannot be described by just a lower bound and an upper bound. Instead, we will only compute the central values for the 11 undetermined master integrals. Following the prescription laid out in Section \ref{sec:bubble}, the central values are defined to maximize the lowest eigenvalue of the matrix $\mathbb M$ in Eq.~\eqref{eq:bananaBlockMatPos}. With the largest cutoff degree 6, there are ${{6+4} \choose 4} = 210$ free parameters in $\vec \alpha$. Therefore, each of the four diagonal blocks in Eq.~\eqref{eq:bananaBlockMatPos} has size $210 \times 210$, and the full matrix $\mathbb M$ has size $840 \times 840$. We use SDPA-QD as the semidefinite programming solver working at quadruple-double precision. The solver is able to take advantage of the block diagonal structure of the matrix $\mathbb M$ to improve efficiency.

\begin{figure}
  \centering
  \includegraphics[width=\textwidth]{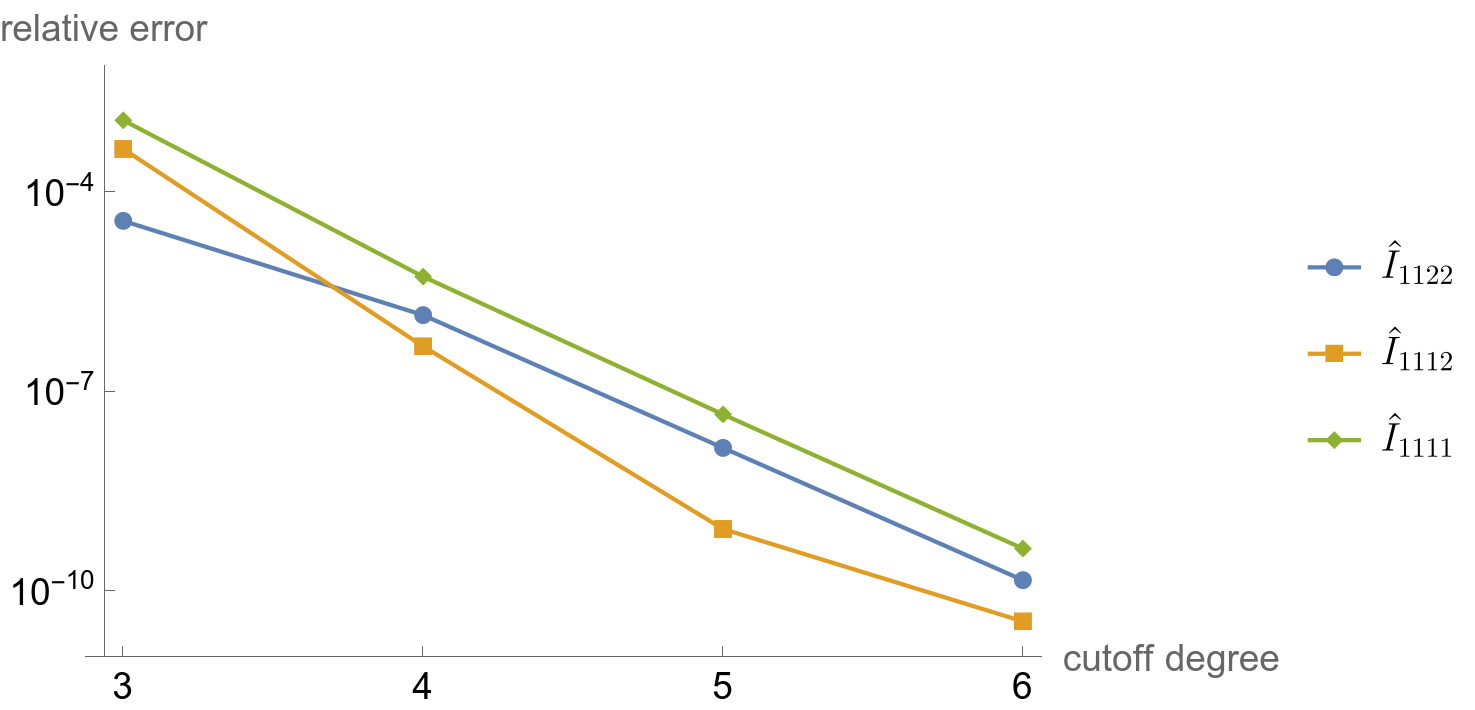}
  \caption{Relative errors of the central values of three representative master integrals of the banana family, versus the cutoff degree in the calculation.}
  \label{fig:bananaErrorVsDeg}
\end{figure}

In Fig.~\ref{fig:bananaErrorVsDeg}, we plot the relative errors of the central values of three representative master integrals against the cutoff degree, for the $\mathcal O(\epsilon^0)$ order only. The actual results are given later in Eq.~\eqref{eq:bananaSampleResults} together with further terms in the $\epsilon$ expansion.
The vertical axis of the plot is on a logarithmic scale, and we can see that the results converge rapidly, in a apparently exponential fashion, as the cutoff degree is raised. With the largest cutoff degree 6, each of the 11 master integrals is evaluated to an accuracy of at least $10^{-9}$.

For the values of master integrals at $\mathcal O(\epsilon^1)$. We will need the result for the parameter values Eq.~\eqref{eq:bananaNumericPoint},
\begin{equation}
  \max \left( \mathcal U^4 / \mathcal F^3 \right) \approx 5000/229059 \, ,
  \label{eq:bananaUFratioMax}
\end{equation}
which we found by numerical maximization in Mathematica. To be conservative, we have slighted rounded up the numerical result to a larger nearby rational number to guarantee that the inequality Eq.~\eqref{eq:bananaMaxUFBlock} is true when using Eq.~\eqref{eq:bananaUFratioMax}. This maximum value occurs at
\begin{equation}
  x_1 \approx 0.12222, \quad x_2 \approx 0.22592, \quad x_3 \approx 0.26701, \quad x_4 \approx 0.38485 \, .
  \label{eq:bananaUFratioMaxPoint}
\end{equation}
While Eq.~\eqref{eq:bananaUFratioMax} will be used directly in calculations, Eq.~\eqref{eq:bananaUFratioMaxPoint} is only included for completeness. The normalization in Eq.~\eqref{eq:bananaUFratioMaxPoint} does not matter since $\mathcal U^4 / \mathcal F^3$ is invariant under the scaling transformation Eq.~\eqref{eq:projective}.

Then the calculation is similar to the calculation of one-loop bubble interals to $\mathcal O(\epsilon^1)$ in Section \ref{sec:bubbleResultsEps}. We use the positivity constraint Eq.~\eqref{eq:bananaPosEps} with Eq.~\eqref{eq:QepsChoice2} for the positive polynomial $Q$, taking the values $k=1,2,3,4$ and combining the constraints from the four different choices. For the $P$ polynomial in Eq.~\eqref{eq:QepsChoice2}, we use a maximum total degree $N_1=6$ for the $\tilde x_i$ variables and a maximum degree of $0$ for the logarithm, i.e. dropping any terms involving the logarithm. The logarithm still appears in the bracket preceding $P^2$ in Eq.~\eqref{eq:QepsChoice2} and contributes to $O(\epsilon)$ parts of integrals by Eq.~\eqref{eq:feynParamRescaledXiEps}. Using the $\mathcal O(\epsilon^0)$ results as known inputs, we again solve a semidefinite programming problem involving a $840 \times 840$ matrix with four diagonal blocks, each of size $210\ \times 210$, to obtain values for the $\mathcal O(\epsilon^1)$ terms of the master integrals.

For the values of master integrals at $\mathcal O(\epsilon^2)$, we use the positivity constraint Eq.~\eqref{eq:bananaPosEps} with Eq.~\eqref{eq:QepsChoice1} for the positive polynomial $Q$. We again use a maximum total degree $N_1=6$ for the $\tilde x_i$ variables but now uses a maximum degree of $1$ for the logarithm. Since the logarithm can appear in a monomial in $P$ with either power 0 or power 1, the size of the matrix in the semidefinite programming problem is doubled to $1680 \times 1680$, with four diagonal blocks each of size $420 \times 420$. Taking both $\mathcal O(\epsilon^0)$ and $\mathcal O(\epsilon^1)$ results as known inputs, we run SDPA-QD to find the central values for the $\mathcal O(\epsilon^2)$ results. For brevity, we show results for 3 representative master integrals out of the 11 top-level master integrals, with kinematic variables taking values of Eq.~\eqref{eq:bananaNumericPoint},
\begin{equation}
  \begin{aligned}
    \hat I_{1122} & \approx 0.31328353052153 -0.12137516161264 \epsilon -1.5577062442336 \epsilon^2, \\
    \hat I_{1112} & \approx 1.3758733318476 -3.5451169250640 \epsilon +0.61363537070259 \epsilon^2, \\
    \hat I_{1111} & \approx 5.9437542439912 -33.914772364319 \epsilon +106.87640125797 \epsilon^2\, .
  \end{aligned}
    \label{eq:bananaSampleResults}  
\end{equation}
For documenting the computational outputs, we have kept each number to 14 significant figures, even their actual accuracies are lower as shown in plots in this section.

We have also calculated both $\mathcal O(\epsilon^1)$ and $\mathcal O(\epsilon^2)$ results using numerical differentiation of integrals evaluated at fixed values of dimensions, following the same strategy of Section \ref{subsec:numericalDiff} for one-loop bubble integrals. The calculations are identical to the $d=2$, i.e.\ $\mathcal O(\epsilon^0)$ case and are based on Eq.~\eqref{eq:bananaPos} without any Taylor expansion in $\epsilon$, with the only change being that $\epsilon$ is set to small numerical values different from 0, i.e.\ $d$ is set to numerical values that slightly deviate from $2$. The 4th-order numerical differentiation formulas, Eqs.~\eqref{eq:finiteDiff1} and Eq.~\eqref{eq:finiteDiff2} are applied with $\epsilon_0=0$ and $\Delta \epsilon = 10^{-3}$, with the spacetime dimension $d=2-2\epsilon$. Example results from this alternative method are, again keeping each number of 14 significant figures,
\begin{equation}
  \begin{aligned}
    \hat I_{1122} & \approx 0.31328353052153 - 0.12137519105424 \epsilon -1.5576503067221 \epsilon^2, \\
    \hat I_{1112} & \approx 1.3758733318476 -3.5451170400199 \epsilon + 0.61369255775305 \epsilon^2, \\
    \hat I_{1111} & \approx 5.9437542439912 -33.914771261794 \epsilon + 106.87318272740 \epsilon^2\, .
  \end{aligned}
    \label{eq:bananaFiniteDiffResults}  
\end{equation}
Note that the $\mathcal O(\epsilon^0)$ results are copied from Eq.~\eqref{eq:bananaSampleResults} as they are not re-calculated.

\begin{figure}[h]
  \centering
  \includegraphics[width=\textwidth]{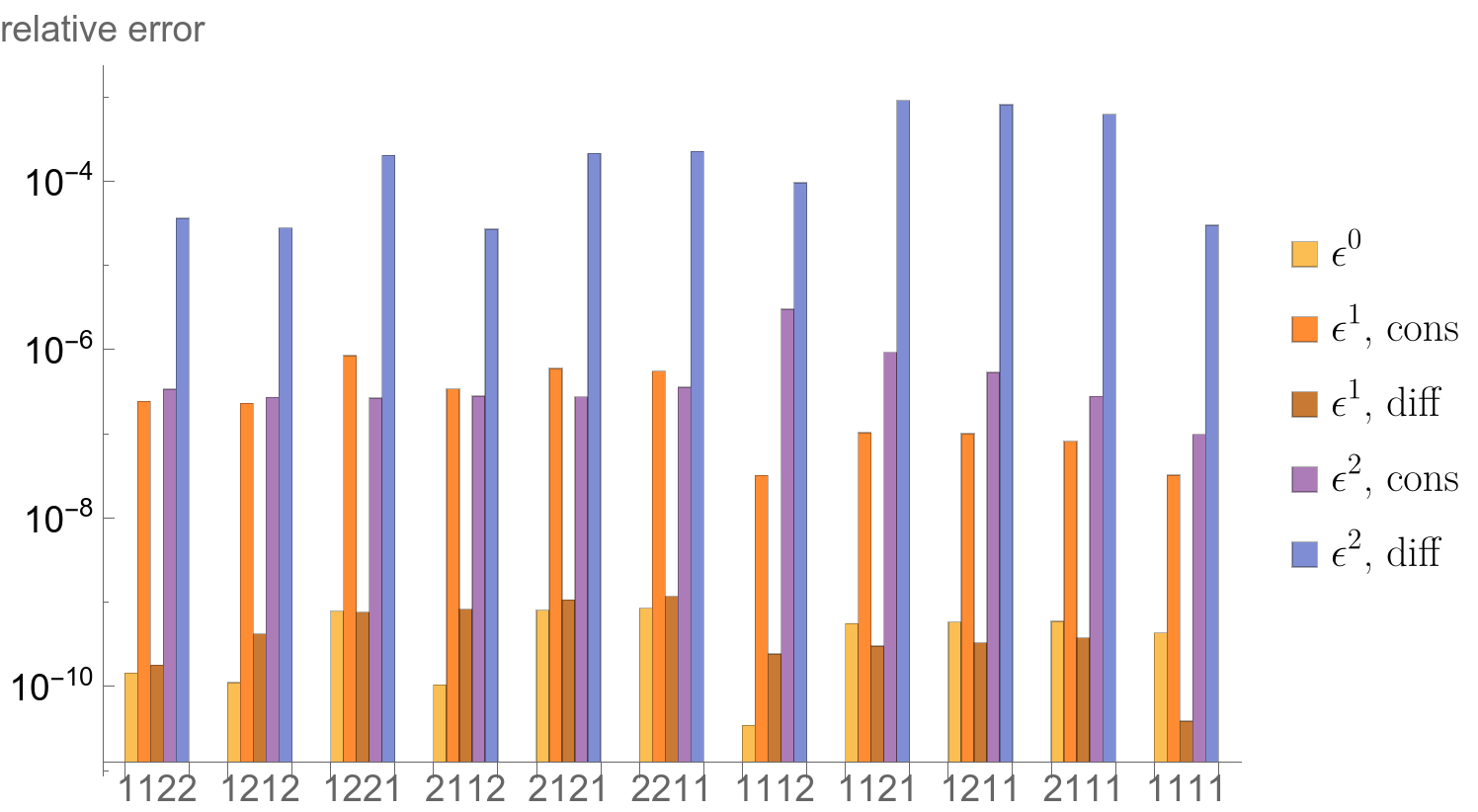}
  \caption{Log-scale plot of relative errors of numerical results for 11 nontrivial master integrals of the three-loop banana family, Eq.~\eqref{eq:mastersBananaTop}, with the normalization Eq.~\eqref{eq:bananaHat}, up to second order in the dimension regularization parameter $\epsilon$. The four numbers under each bar indicates the subscript indices of $\hat I_{a_1, a_2, a_3, a_4}$, with commas omitted as all four indices are single-digit numbers (either 1 or 2). For each master integral, the five vertical bars, from left to right, show the relative errors for the $\epsilon^0$ term, the $\epsilon^1$ term calculated from direct positivity constraints (abbreviated as {\tt cons} in the legend), the $\epsilon^1$ term calculated from numerical differentiation (abbreviated as {\tt diff} in the legend), the $\epsilon^2$ term calculated from direct positivity constraints, and the $\epsilon^2$ term calculated from numerical differentiation.}
  \label{fig:bananaCombinedChart}
\end{figure}
These numerical results for the $\epsilon$ expansion of master integrals Eq.~\eqref{eq:mastersBananaTop} are obtained with the normalization of Eq.~\eqref{eq:bananaHat}. The reference results from DiffExp are have the normalization of Eq.~\eqref{eq:banana} and additional factors for individual master integrals. The reference results have been converted to use our normalizations for comparison. DiffExp results for the three sample integrals, truncated to 14 significant digits, are
\begin{equation}
    \begin{aligned}
      \hat I_{1122} & \approx 0.31328353056677-0.121375191032390 \epsilon-1.5577067713048 \epsilon^2 , \\
      \hat I_{1112} & \approx 1.37587333189510-3.5451170391547 \epsilon+0.61363351857945 \epsilon^2, \\
      \hat I_{1111} & \approx 5.9437542414259-33.914771263107 \epsilon+106.876390717227 \epsilon^2 \, .
    \end{aligned}
    \label{eq:bananaDiffExpResults}  
\end{equation}
The final numerical accuracy for the 11 master integrals, with values of kinematic parameters chosen in Eq.~\eqref{eq:bananaNumericPoint}, is shown in Fig.~\ref{fig:bananaCombinedChart}.
The $\epsilon$ expansion results from ``direct positivity constraints'' are labeled {\tt cons} and results from numerical differentiation are labeled {\tt diff} in the plot legend. We can see that numerical differentiation gives very good accuracy for $\mathcal O(\epsilon^1)$ terms, comparable with the accuracy of $\mathcal O(\epsilon^0)$ terms, while for the $\mathcal (\epsilon^2)$ terms, direct positivity constraints yield more accurate results. In any case, both methods for the $\epsilon$ expansion have demonstrated their potentials in this initial investigation, as all results for $\mathcal O(\epsilon^1)$ terms have relative errors below $10^{-6}$ and all results for $\mathcal O(\epsilon^2)$ terms have relative errors below $10^{-3}$.

We now comment on computational resources used, when using the highest cutoff degree 6. The computation times below are obtained without using multiple CPUs. IBP reduction takes a few hours with FIRE6 \cite{Smirnov:2019qkx} with numerical kinematics Eq.~\eqref{eq:bananaNumericPoint}. The IBP reduction results are obtained with analytic dependence on $d$ and can be subsequently expanded in $\epsilon$, so no extra IBP reduction is needed for obtaining the $\epsilon$ expansions of master integrals beyond the zeroth order. Running the semidefinite programming solver SDPA-QD takes a few hours for every run, including one run for solving positivity constraints for the $\mathcal O(\epsilon^i)$ part for each $i=0, 1, 2$, and for the alternative method based on numerical differentiation, several runs at different numerical values of $\epsilon$ to generate the data needed to feed into finite-difference approximations.

\begin{figure}
  \centering
  \includegraphics[width=.8\textwidth]{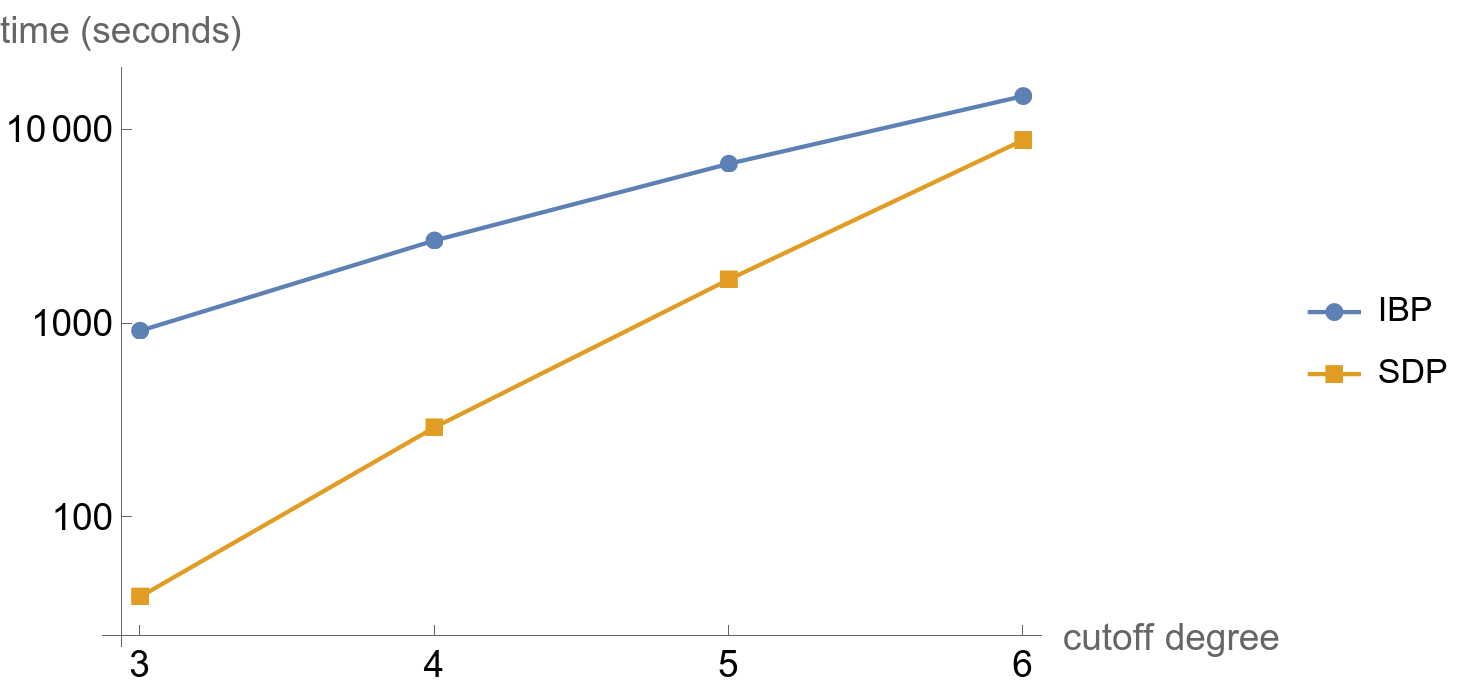}
  \caption{Computation time taken to evaluate the master integrals at order $\epsilon^0$, versus the cutoff degree. IBP reduction (FIRE6) and semidefinite program solving (SDPA-QD) times are shown separately.}
  \label{fig:timingVsDeg}
\end{figure}

Furthermore, Fig.~\ref{fig:timingVsDeg} shows how the computation time depends on the cutoff degree used for computing the $\epsilon^0$ part of the master integrals. This is of interest since the cutoff degree in turn controls the accuracy reached, c.f.~Fig.~\ref{fig:bananaErrorVsDeg}). The computation time is broken up into IBP reduction (with FIRE6) and the solving of the semidefinite program (with SDPA-QD).\footnote{For practical reasons, two different server machines were used. FIRE6 was run on a machine with Intel Xeon E5-2697A v4, and SDPA-QD was run on a machine with Intel Xeon Gold 6130. While crude testing indicates that the performance of the two machines are comparable, the essential information is in the scaling behavior of each individual curve.}
The size of each of the four diagonal blocks of the matrix Eq.~\eqref{eq:bananaBlockMatPos} is equal to 
$\binom{4+k}{4}$ for a cutoff degree of $k$, due to the fact that there are $4$ propagators in the three-loop banana diagram. For $k=3, 4, 5, 6$, the sizes are $35, 70, 126, 210$, respectively. The time spent on solving the semidefinite program is very well described a cubic scaling with the matrix size, while the IBP reduction time has a less steep growth as the cutoff degree is raised.
%Therefore, in the limit of a high cutoff $k$, the computation time is expected to scale like $\binom{4+k}{4}^3 \sim k^{12}$.

To put the timing into context, we compare with numerical series solutions of differential equations with the DiffExp package, which generated the reference results in this study. Constructing the differential equations via IBP reduction (with FIRE6) subject to kinematics Eq.~\eqref{eq:bananaDELine} and solving the differential equations (with DiffExp) took a total of less than 18 minutes on a less powerful laptop computer, even though roughly five times as many significant digits are reached compared with our results based on positivity constraints. Therefore, for this problem, our method is not competitive against differential equations. However, it is worth noting that our method is in principle applicable to single-scale integrals, such as vacuum integrals with uniform internal masses, without introducing additional scales. Also, in more general cases, constructing differential equations requires IBP reduction with analytic dependence on at least one kinematic variables, while our method requires IBP reduction at one specific point in the space of kinematic variables, even though more complicated integrals (e.g.\ with higher propagator powers) will need to be reduced. Therefore, numerical and finite-field IBP reduction techniques (see e.g.\ \cite{vonManteuffel:2014ixa, Peraro:2016wsq, Peraro:2019svx, Klappert:2019emp}) can be potentially well suited for speeding up our computations in future studies.

\section{Discussions}
\label{sec:discussions}
We have demonstrated a new method for evaluating Feynman integrals which, to our best knowledge, is the first method based on inequality constraints, while previously exploited consistency conditions for Feynman integrals are based on equality constraints (such as the vanishing of the coefficient of a certain spurious singularity). Our calculation strategy is writing down an infinite class of convergent integrals with non-negative integrands and reducing them to linear sums of a set of master integrals. This constraints an infinite number of linear sums of master integrals to be non-negative. A truncated set of the constraints can be solved as a semidefinite programming problem in mathematical optimization. Surprisingly, the constraints appear strong enough to determine the integrals to any desired precision since the bounds appear to converge exponentially as the truncation cutoff is increased. Like the method of differential equations \cite{Kotikov:1990kg,Bern:1993kr,Remiddi:1997ny, Gehrmann:1999as,Henn:2013pwa}, our method relies on integration-by-parts (IBP) identities, but instead of using differential equations to transport the values of the integrals across kinematic space, we only use IBP information at a single point in kinematic space.

Though our study is preliminary, the numerical results are promising. We have demonstrated the applicability of our methods to a nontrivial example, namely three-loop banana integrals with four unequal internal masses in $d= 2 - 2\epsilon$ dimensions. With modest computational resources, we evaluated the $\mathcal O(\epsilon^0)$ part of all the 11 nontrivial master integrals to a relative accuracy of at least $10^{-9}$. The accuracies for $\mathcal O(\epsilon^1)$ and $\mathcal O(\epsilon^2)$ terms are lower, though only slightly so for $\mathcal O(\epsilon^1)$ terms when the numerical differentiation method is used. For all but the smallest problems, extended-precision floating point arithmetic is needed to ensure numerical stability in the semidefinite programming solver, similar to what was encountered in the conformal bootstrap \cite{Simmons-Duffin:2015qma} and the quantum mechanics bootstrap \cite{Berenstein:2022unr}. We note that extended precision is also generally need in the evaluation of Feynman integrals by series solutions of differential equations as observed in e.g.\ Refs.~\cite{Moriello:2019yhu, Hidding:2020ytt}.

We have also revealed hidden consistency relations that link different terms in the $\epsilon$ expansions of Feynman integrals. As explained in Section \ref{subsubsec:eps_generic}, for any (quasi-) finite Feynman integral without numerators, the $\epsilon$ expansion terms (appropriately normalized) must give rise to a positive-semidefinite Hankel matrix. This is an extremely general statement which can be checked against a huge number of Feynman integral computations in the literature, because many Feynman integrals have Euclidean regions and it is believed that a quasi-finite basis exist for any family of integrals \cite{vonManteuffel:2014qoa}. This result is an elementary consequence of our analysis but has not been previously exposed in the literature. Such constraints have been solved numerically in our paper to predict the $\epsilon$ expansion terms to high accuracy. We have also formulated an alternative method to obtain the $\epsilon$ expansion terms by numerical differentiation of semidefinite programming solutions with respect to the spacetime dimension. The above two methods for calculating $\epsilon$ expansion terms are complementary and we have found cases in which either of them outperforms the other in accuracy.

Our new method for calculating Feynman integrals is analogous to recent developments in bootstrapping quantum mechanics systems and lattice models \cite{Lin:2020mme, Han:2020bkb, Berenstein:2021dyf, Berenstein:2022unr, Anderson:2016rcw, Kazakov:2022xuh, Cho:2022lcj}. For example, the role of IBP and dimensional-shifting identities in our work is analogous to the role of moment recursion relations in the quantum mechanics bootstrap. Analogous identities also appear in EFT bounds as ``null constraints'' from crossing symmetry \cite{Caron-Huot:2020cmc}.\footnote{We thank Francesco Riva for pointing out this connection.} While our work has imported techniques developed in non-perturbative contexts to perturbative physics, in the reverse direction, the differential equation method of perturbative calculations has been applied to non-perturbative lattice correlation functions in Refs.~\cite{Gasparotto:2022mmp, Weinzierl:2020nhw, Cacciatori:2022mbi}, also exploiting identities similar to those from IBP. Therefore, we expect a fruitful exchange of techniques between perturbative and non-perturbative calculations.

Finally, we speculate on possible future work. Except for the generic constraints on the $\epsilon$ expansion, this paper has mainly treated massive Feynman integrals, and for integral families involving massless internal lines, it would be necessary to identify non-negative integrals free of not only ultraviolet but also infrared divergences to generate the positivity constraints. To extend our method to integrals outside the Euclidean region, it remains to be seen how positivity constrains can be formulated, possibly for real and imaginary parts separately after a suitable deformation of the integration contour. Connections with other notions of positivity relevant for Feynman integrals \cite{Abreu:2019wzk, Yang:2022gko} remain to be explored. Another interesting question is whether positivity constraints can be used to understand complete scattering amplitudes (rather than individual Feynman integrals) at a fixed order in perturbation theory, in light of numerical observations in $\mathcal N=4$ super-Yang-Mills theory amplitudes in Ref.~\cite{Dixon:2016apl}.

\section*{Acknowledgments}
M.Z.’s work is supported in part by the U.K.\ Royal Society through Grant URF\textbackslash R1\textbackslash 20109. We thank York Schroeder for useful comments on the manuscript. For the purpose of open access, the author has applied a Creative Commons Attribution (CC BY) license to any Author Accepted Manuscript version arising from this submission.

\appendix
\section{Analytic results for bubble integrals}
\label{app:bubbleAnalytic}
The material here is well known in the literature but included for completeness. We work in spacetime dimension $d=4-2\epsilon$. Using Feynman parametrization, Eq.~\eqref{eq:selfEnergyInt} is written as, for $a_1 \geq 1, a_2 \geq 1$.
\begin{equation}
  I^d_{a_1, a_2} = \frac {\Gamma(a_1 + a_2 - d / 2) e^{\gamma_E \epsilon}}
  {\Gamma(a_1) \Gamma(a_2)} \int_0^1 dx \, x^{a_1 - 1} (1-x)^{a_2 - 1}
  \left[ m^2 - p^2 x (1-x) - i 0^+ \right]^{d/2 - a_1 - a_2} \, .
\end{equation}
This allows us to evaluate the first of the master integrals in Eq.~\eqref{eq:bubbleFiniteBasis} as
\begin{equation}
  I^{d=4-2\epsilon}_{2,1} = \Gamma(1 + \epsilon) (m^2)^{-1-\epsilon} e^{\gamma_E \epsilon} \int_0^1 dx \, \frac x {\left[1 - x (1-x)  p^2/m^2 - i0^+ \right]^{1 + \epsilon}}
\end{equation}
Since the denominator in the above integrand is symmetric under $x \leftrightarrow 1-x$, we can symmetrize the numerator as $x \rightarrow [x + (1-x)] / 2 = 1/2$, obtaining
\begin{equation}
  I^{d=4-2\epsilon}_{2,1} = \Gamma(1+\epsilon) (m^2)^{-1-\epsilon} e^{\gamma_E \epsilon} \int_0^1 dx \, \frac 1 {2 \left[1 - x (1-x) p^2 / m^2 -i0^+ \right]^{1 + \epsilon}}
  \label{eq:bubbleAnalytic}
\end{equation}
For $p^2 < 0$, this evaluates to
\begin{equation}
  I^{d=4-2\epsilon}_{2,1} = \Gamma(1+\epsilon) (-p^2 + i0^+)^{-1-\epsilon} e^{\gamma_E \epsilon} \left[ \frac 1 \beta \log \frac {\beta+1} {\beta-1} + \mathcal O(\epsilon) \right] \, ,
\end{equation}
where we defined
\begin{equation}
  \beta = \sqrt{ 1- \frac{4m^2}{p^2} - i0^+ } \, . \label{eq:bubbleBeta}
\end{equation}
The omitted $\mathcal O(\epsilon)$ term in Eq.~\eqref{eq:bubbleAnalytic} is easy to evaluate and we do not include the explicit result here. The second master integral in Eq.~\eqref{eq:bubbleFiniteBasis} is a trivial tadpole integral,
\begin{equation}
  I^{d=4-2\epsilon}_{3,0} = \frac 1 2 \Gamma(3 - d/2) e^{\gamma_E (4-d)/2} (m^2)^{-3+d/2} = \frac 1 2 \Gamma(1 + \epsilon) e^{\gamma_E \epsilon} (m^2)^{-1-\epsilon} \, . \label{eq:tadpoleAnalytic}
\end{equation}
So the ratio defined in Eq.~\eqref{eq:bubbleMasterParam} is equal to, when $d=4-2\epsilon$,
\begin{align}
  \hat I^{d=4-2\epsilon}_{2,1} &= I^{d=4-2\epsilon}_{2,1} / I^{d=4-2\epsilon}_{3,0} = 2 \int_0^1 dx \, \frac x {\left[1 - x (1-x)  p^2/m^2 - i0^+ \right]^{1 + \epsilon}} \nonumber \\
  &= -\frac {2m^2}{p^2} \frac 1 \beta \log \frac {\beta+1} {\beta-1} + \mathcal O(\epsilon) \, .
\end{align}
This result is real and positive for any $p^2 < 4 m^2$, after taking into cancellation of imaginary parts with the definition Eq.~\eqref{eq:bubbleBeta}. The corrections at higher order in the dimensional regularization parameter $\epsilon$ are also readily obtained but we do not include the results here.

The tadpole integral with an arbitrary power for the propagator is well known,
\begin{equation}
  \int \frac{d^d l \, e^{\gamma_E \epsilon}} {i \pi^{d/2}} \frac 1 {(-l^2+m^2)^n} =
  \frac 1 2 e^{\gamma_E (4-d)/2} \frac {\Gamma(n - d/2)} {\Gamma(n)} \frac 1 {(m^2)^{n - d/2}} \, . \label{eq:tadpoleGeneral}
\end{equation}

%\bibliography{refs.bib, more_refs.bib}

\providecommand{\href}[2]{#2}\begingroup\raggedright\endgroup

\end{document}